\documentclass[sigconf]{acmart}
\usepackage{lipsum}
\usepackage{todonotes}
\usepackage{xcolor}
\usepackage{algorithm,algpseudocode}
\usepackage{amsmath}
\usepackage{bbm}
\usepackage{multirow}
\usepackage{colortbl}
\usepackage{makecell}
\usepackage{acmart-taps}

\newlength\myheight
\newlength\mydepth
\newcommand*\inlinegraphics[1]{%
  \settototalheight\myheight{Xygp}%
  \settodepth\mydepth{Xygp}%
  \raisebox{-3pt}[0pt][-10pt]{\includegraphics[height=1.5\myheight]{#1}}%
}
\newcommand*\inlinegraphicsSmall[1]{%
  \settototalheight\myheight{Xygp}%
  \settodepth\mydepth{Xygp}%
  \raisebox{-1pt}[0pt][-10pt]{\includegraphics[height=\myheight]{#1}}%
}

\definecolor{DiverseMagenta}{rgb}{0.75, 0.0, 0.75}
\definecolor{AccentBlue}{rgb}{0.0, 0.5, 1.0}

\definecolor{mintgreen}{RGB}{152, 255, 152}
\makeatletter
\newcommand*\iftodonotes{\if@todonotes@disabled\expandafter\@secondoftwo\else\expandafter\@firstoftwo\fi}  %
\makeatother

\usepackage{xcolor , acmart-taps}
\usepackage{soul}  
\usepackage{xifthen}

\newboolean{showrevisions}
\setboolean{showrevisions}{true} 

\definecolor{addcolor}{RGB}{0,100,200}     
\definecolor{delcolor}{RGB}{200,0,0}       

\definecolor{red}{RGB}{224,102,102}
\AtBeginDocument{%
  }

\setcopyright{none}
\acmConference{}{}{}
\acmBooktitle{}
\acmDOI{}
\acmISBN{}

\renewcommand\footnotetextcopyrightpermission[1]{}

\setcopyright{none}

\aptLtoXcmd{\long\def\A#1{\xbox{aptbox}{\XMLaddatt{style}{width: 900px;  background-color: \#c7c7ff; border: 1px solid \#00008c; padding: 10px;}{#1}}}}{}

\aptLtoXcmd{\long\def\I#1{\xbox{aptbox}{\XMLaddatt{style}{width: 900px;  background-color: \#ffe3c7; border: 1px solid \#aa6f35; padding: 10px;}{#1}}}}{}

\aptLtoXcmd{}{
\definecolor{customgreen}{HTML}{93C47D}
\def\yes{\sq{customgreen}}
\definecolor{customred}{HTML}{E06666}
\def\no{\sq{customred}}
\definecolor{customyellow}{HTML}{FFD966}
\def\partialminor{\sq{customyellow}}
\definecolor{customorange}{HTML}{F6A623}
\def\partialmajor{\sq{customorange}}
\newcommand{\badge}[2]{%
  \tikz[baseline=(X.base)]\node[inner xsep=3pt,inner ysep=1pt,rounded corners=2pt,
    fill=#2!22, draw=#2!55!black] (X) {\scriptsize\textsf{#1}};}

\newcommand{\I}{\badge{I}{orange}}
\newcommand{\A}{\badge{A}{blue}}}

\definecolor{green}{RGB}{147,196,125}

\definecolor{color1}{RGB}{255 , 217 , 102}

\definecolor{color2}{RGB}{246,166,35}

\aptLtoXcmd{\def\no{{\setlength{\fboxsep}{0pt}\colorbox{red}{\rule{0pt}{6pt}\hspace{10pt}}}}}{}
\aptLtoXcmd{\def\yes{{\setlength{\fboxsep}{0pt}\colorbox{green}{\rule{0pt}{6pt}\hspace{10pt}}}}}{}

\aptLtoXcmd{\def\partialminor{{\setlength{\fboxsep}{0pt}\colorbox{color1}{\rule{0pt}{6pt}\hspace{10pt}}}}}{}
\aptLtoXcmd{\def\partialmajor{{\setlength{\fboxsep}{0pt}\colorbox{color2}{\rule{0pt}{6pt}\hspace{10pt}}}}}{}
\begin{document}

\title{\textsc{PleaSQLarify}: Visual Pragmatic Repair for Natural Language Database Querying}

\author{Robin Shing Moon Chan}
\orcid{0009-0007-1319-8334}
\affiliation{%
  \institution{ETH Zurich}
  \city{Zurich}
  \country{Switzerland}
}
\email{robin.chan@inf.ethz.ch}

\author{Rita Sevastjanova}
\orcid{0000-0002-2629-9579}
\affiliation{%
  \institution{ETH Zurich}
  \city{Zurich}
  \country{Switzerland}}
\email{rita.sevastjanova@inf.ethz.ch}

\author{Mennatallah El-Assady}
\orcid{0000-0001-8526-2613}
\affiliation{%
  \institution{ETH Zurich}
  \city{Zurich}
  \country{Switzerland}
}
\email{melassady@ai.ethz.ch}

\renewcommand{\shortauthors}{Chan et al.}

\begin{abstract}
    Natural language database interfaces broaden data access, yet they remain brittle under input ambiguity. Standard approaches often collapse uncertainty into a single query, offering little support for mismatches between user intent and system interpretation. We reframe this challenge through pragmatic inference: while users economize expressions, systems operate on priors over the action space that may not align with the users'. In this view, pragmatic repair---incremental clarification through minimal interaction---is a natural strategy for resolving underspecification. We present \textsc{PleaSQLarify}, which operationalizes pragmatic repair by structuring interaction around interpretable decision variables that enable efficient clarification\footnote{Code: \href{https://github.com/chanr0/pleasqlarify}{github.com/chanr0/pleasqlarify}}. A visual interface
     complements this by surfacing the action space for exploration, requesting user disambiguation, and making belief updates traceable across turns. In a study with twelve participants, \textsc{PleaSQLarify} helped users recognize alternative interpretations and efficiently resolve ambiguity. Our findings highlight pragmatic repair as a design principle that fosters effective user control in natural language interfaces.
\end{abstract}


\begin{teaserfigure}
  \includegraphics[width=\textwidth]{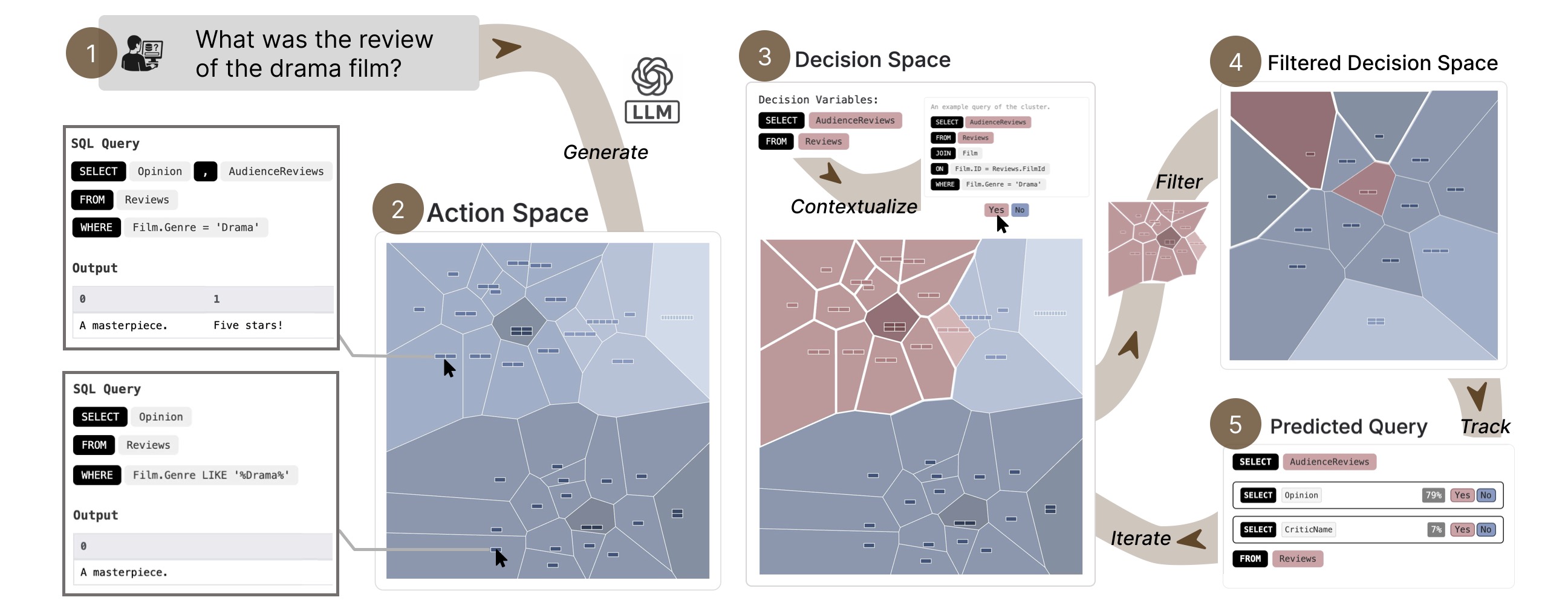}
  \vspace{-2em}
  \caption{Overview of \textsc{PleaSQLarify}, an interactive approach to pragmatic repair applied to text-to-SQL. (1) A user poses an underspecified natural language query. (2) The system generates an initial action space of possible SQL interpretations. (3) The user is presented with the most \textit{informative} decision variable, i.e., one that splits the action space in a meaningful way. (4) The user iteratively filters the space by engaging with clarifications that contrast alternative interpretations. (5) The predicted query is continuously updated and made traceable across turns, enabling the user to explore, compare, and refine interpretations until their intent is matched.}
  \Description{Diagram with five numbered panels connected by arrows. Panel 1 shows a user query (“What was the review of the drama film?”) leading to an SQL query and output. Panel 2 displays a mosaic-like “Action Space” of many small cells representing query variations. Panel 3 highlights decision variables in a color-coded map labeled “Decision Space.” Panel 4 shows a filtered version of this map, narrowing down options. Panel 5 displays the final “Predicted Query” box with tracked SQL components. Arrows loop between panels 3 and 4, indicating iterative refinement, while the flow moves from 1 through 5 overall.}
  \label{fig:teaser}
\end{teaserfigure}

\maketitle

\pagestyle{plain}

\section{Introduction}

Natural  language interfaces to data systems allow users to query and interact with databases using everyday language \cite{gao2015datatone, setlur2019inferencing, shaw-etal-2021-compositional, gao2024text2sql}.
For instance, a data analyst might type "\textit{show me last quarter's top customers}" into a natural language database interface to explore an unfamiliar dataset.
These interactions feel natural because they allow people to express intent in their own words, without needing technical commands or formal syntax.
Yet the same naturalness also makes them fragile: such requests are often ambiguous, and systems typically sample the most probable interpretation and perform the corresponding system action \cite{luger2016pa, gao2015datatone, smith2012helping}.
This challenge is particularly prominent in natural language programming, such as text-to-SQL \cite{shaw-etal-2021-compositional, gao2024text2sql, zhong2017seq2sqlgeneratingstructuredqueries, lawrence-riezler-2018-improving}, where natural utterances must be translated into precise database queries.
Especially in such tasks, seemingly straightforward requests often have more than one valid interpretation \cite{saparina2024ambrosia}. Evidence from the CoSQL dataset \cite{yu2019cosql} underscores this point: across the 2'459 interactions where users asked SQL experts to write basic SQL queries conversationally, 31.9\% of which contained an instruction requiring clarification of the user's intent. Although some systems attempt to manage such ambiguity through interactive interfaces \cite{liu2025nli4db, narechania2021diy}, current systems still do not expose it effectively or provide robust guiding mechanisms for clarification.\looseness=-1

Taking a step back, this problem reflects a basic property of language use: \textit{underspecification}.
Pragmatic theories of language use \cite{grice1975logic, piantadosi2012communicative, frank2012rsa} describe humans as economical speakers who deliberately leave details unspecified, relying on listeners to recover the intended meaning from context and shared knowledge. 
Such underspecification is a natural outcome of human communication; however, in natural language interfaces, it leaves systems with multiple plausible actions and users unaware of which ones the system considered or why.
Akin to clarification in human communication, we propose that ambiguity should be interactively engaged with directly through \textit{pragmatic repair}.
Namely, in human conversation, underspecified expressions are routinely clarified incrementally: speakers add just enough information to distinguish their intent from alternatives, and listeners offer targeted requests for clarification. By analogy, natural language interfaces may resolve ambiguity more effectively by surfacing alternative interpretations and asking for clarifications that are both efficient and intelligible to users. 

To operationalize this idea, we first propose a general framing of pragmatic repair in natural language interfaces, which we then instantiate for text-to-SQL, presenting \textsc{PleaSQLarify}. 
As showcased in~Figure~\ref{fig:teaser}, the approach combines an algorithm that selects informative, grouped decision variables with a visual interface that (1) surfaces the model’s action space, (2) supports guided narrowing through contextualized clarifications, and (3) makes updates traceable across turns. Rather than forcing users to accept a single opaque query, the system invites them to explore, compare, and refine interpretations until a satisfactory query is reached.
In a study with twelve participants, \textsc{PleaSQLarify} helped users recognize alternative interpretations they had not anticipated and efficiently steer toward satisfactory results. Interaction traces revealed three utterance clarification workflows---broad exploration, targeted prefiltering, and output-driven assembly---that illustrate how users flexibly adapted repair strategies to their goals and the complexity of the task.
Overall, this paper makes four contributions:
\begin{itemize}
\item A conceptual framing of pragmatic repair for natural language interfaces. We extend theories of pragmatic inference to HCI, showing how underspecification in user utterances can be used as a resource for incremental clarification.
\item An algorithm for pragmatic repair in natural language interfaces, and an instantiation for text-to-SQL. The algorithm groups atomic features into semantically interpretable decision variables and selects among them by expected information gain, enabling clarification that is both efficient and meaningful. A quantitative evaluation on the AMBROSIA dataset shows that this approach reduces semantic uncertainty in fewer turns than baselines.
\item A visual interface that operationalizes pragmatic repair for text-to-SQL by making the model’s action space visible, allowing users to surface informative decision variables, and maintaining traceability of updates across turns.
\item Empirical insights from a user study that show the effectiveness of the proposed solution and design implications for supporting repair in natural language interfaces more broadly.\looseness=-1
\end{itemize}

\section{Related Work}\label{sec:related-work}
This section surveys the current state of research on the concept of ambiguity in text-to-SQL, as well as algorithm- and user feedback-based disambiguation approaches.

\paragraph{Ambiguity in Text-to-SQL} 
A recent survey by Liu et al.~\cite{liu2025nli4db} identifies ambiguity as a critical unresolved challenge in natural language interfaces for databases. 
Various forms of ambiguity, such as scope and vagueness, have been identified by researchers, prompting the creation of taxonomies to organize them systematically (e.g.,~\cite{chen-etal-2023-text,floratou2024nl2sql}).
Several benchmarks have emerged to address the different types of ambiguity, including the CoSQL dataset~\cite{yu2019cosql}, which contains over 10'000 annotated SQL queries with clarified ambiguities and flagged unanswerable questions; SPLASH~\cite{elgohary-etal-2020-speak}, which contains utterances with incorrect SQL interpretations and natural language feedback; AmbiQT~\cite{bhaskar2023benchmarking}, which was created using a novel decoding algorithm that combines plan-based template generation with constrained infilling; and AMBROSIA~\cite{saparina2024ambrosia}, a benchmark that includes three distinct types of ambiguity: scope ambiguity, attachment ambiguity, and vagueness, with queries designed to support multiple valid interpretations reflecting real-world complexity. 
In this work, we employ the AMBROSIA dataset both for the quantitative evaluation of our proposed algorithm and for the extraction of query samples used in the user study.\looseness=-1

\paragraph{Algorithm-Based Disambiguation}
In recent years, several algorithmic approaches have been proposed for natural language query disambiguation used at different stages of the text-to-SQL pipeline.
For instance, \citet{kim-etal-2024-aligning} introduce APA (Alignment with Perceived Ambiguity) pipeline that aligns LLMs to manage ambiguous queries before their execution.
Another example is the ``Disambiguate First, Parse Later'' paradigm that uses a two-stage approach, initially resolving ambiguity by generating all possible natural language meanings and subsequently parsing each resulting unambiguous interpretation~\cite{saparina2025disambiguate}. 
\citet{wang2023know} propose a weakly supervised model, DTE (Detecting-Then-Explaining), to handle ambiguous and unanswerable questions by framing token localization as a sequence labeling task, tagging each token as ambiguous, unanswerable, or other.
Ambiguity resolution can also be carried out during SQL translation. For instance,~\citet{chen2025reliable} use pre-trained classifiers alongside conformal prediction methods to assess the likelihood of error detection during the translation step.
Recent contemporaneous work by \citet{qiu2025interactive} proposes an SQL disambiguation method that utilizes similarity-agnostic, clause-level expected information gain while neglecting the functional equivalence between queries and the interaction between features.

\paragraph{User Feedback-Based Disambiguation}
Beyond purely algorithmic approaches, interactive visual systems are frequently employed to allow users to address ambiguities through direct manipulation and feedback.
The systems, however, vary in their approaches to providing feedback.
For instance, Debug-It-Yourself (DIY) provides a sandbox environment for users to interact with question-to-query mappings and database subsets, allowing adjustments when errors are detected~\cite{narechania2021diy}. 
An interactive framework called FISQL (Feedback-Infused SQL Generation Tool) allows for the refinement of SQL generation through user-provided natural language feedback by utilizing prompting strategies~\cite{menon2025fisql}. 
AmbiSQL uses in-context learning to detect ambiguous phrases, classify them, and generate targeted clarification questions with multiple-choice options for query rewriting~\cite{ding2025ambisql}. 
Several works focus on disambiguation within specific domains. 
For instance, DataTone employs mixed-initiative disambiguation to turn natural language queries into visualizations using interactive ambiguity widgets that surface system decisions and allow immediate user resolution~\cite{gao2015datatone}. 
User corrections are stored as constraints that influence subsequent queries, creating a learning system. 
\citet{setlur2019inferencing} introduce a system to resolve ambiguity in utterances when used in visual analytics interfaces. They use syntactic and semantic constraints along with heuristics to restrict the solution space. 
\citet{ma2025ambigchat} extend hierarchical clarification strategies to open-domain question answering by constructing disambiguation trees that systematically decompose ambiguous queries into navigable interpretation spaces, sharing conceptual similarities with our approach, where schema elements and join paths form natural decision hierarchies.
A more empirical study by \citet{zamfirescu2023johnny} reveals how non-expert users engage in iterative query refinement; strongly preferring low-formulation-cost strategies such as selecting from presented options over writing detailed specifications from the outset.
\citet{poesia2021pragmatic} apply interactive disambiguation to programming languages, allowing programmers to utilize a controlled level of ambiguity.\looseness=-1

Our work builds upon these foundations by focusing on both the algorithmic and interface sides of ambiguity resolution in text-to-SQL systems. 
Our goal is to move beyond the black-box nature of recent approaches that utilize LLMs for disambiguation~\cite{kim-etal-2024-aligning,ding2025ambisql,menon2025fisql,saparina2025disambiguate} towards a human-centered and interpretable framework, where every user decision and their impact can be interpreted.
In particular, we propose a solution that uses pragmatic language theories as a foundation to build an algorithm that selects informative, grouped decision variables and combine the algorithm with visual interactive components for guided and informed query disambiguation.

\section{Background: Pragmatics}
\label{subsec:pragmatics}

\paragraph{Pragmatic Inference and RSA} 
First, let $u \in \mathcal{U}$ be a natural language utterance chosen by a user from the set of possible utterances $\mathcal{U}$ to communicate an intended meaning $m^* \in \mathcal{M}$. In the rational speech act theory (RSA) of probabilistic pragmatics \cite{frank2012rsa}, communicative efficiency is captured as 
\begin{equation}
    p_S(u\mid m^*) \propto \exp(\alpha \cdot U(u, m^*) - C(u)),
\end{equation}
where $U(u, m^*)$ measures how well $u$ distinguishes $m^*$ from alternatives, and $C(u)$ measures the production cost of $u$. We note that $U(u, m)$ is shaped by the speaker's prior over meanings $p_S(m)$. The interpreter's, or listener's $L$, task is to infer a meaning $m$ from the speaker's chosen utterance $u \sim p_S(u\mid m^*)$. In RSA, the meaning is conceptually sampled from a posterior $m \sim p_L(m\mid u)$ defined through Bayes' rule, given the listener's prior over meanings $p_L(m)$, 
\begin{equation}
    p_L (m\mid u) \propto p_S(u\mid m) \cdot p_L(m).
\end{equation}

\paragraph{Pragmatic Repair}
In human–human communication, \textit{utterance grounding} refers to the interactive process by which speaker and listener establish mutual understanding that the utterance has been correctly interpreted in context, thereby aligning their beliefs about $m^*$. If there is shared \textit{common ground} between speaker and listener, their priors are well-aligned. In such cases, even relatively underspecified utterances may yield a low-entropy, peaked posterior $p_L(m\mid u)$.
The pragmatic mechanism through which mutual understanding is maintained is \textit{pragmatic repair}, that is, the incremental clarification that occurs when an utterance is ambiguous or underdetermined.
In Clark and Wilkes-Gibbs’s \cite{clark1986referring} collaborative model of reference, pragmatic repair proceeds incrementally: speakers often begin with a minimal, sometimes general, description of a referent and refine it step-by-step, adding just enough new information to distinguish the intended meaning from remaining alternatives. This coarse-to-fine refinement reflects the pragmatic principle of \textit{least collaborative effort} \cite{grice1975logic}, in which interlocutors avoid over-specification up front and instead contribute additional details only when needed to resolve remaining uncertainty.

\begin{figure*}[t!]
    \centering
    \includegraphics[width=0.9\linewidth]{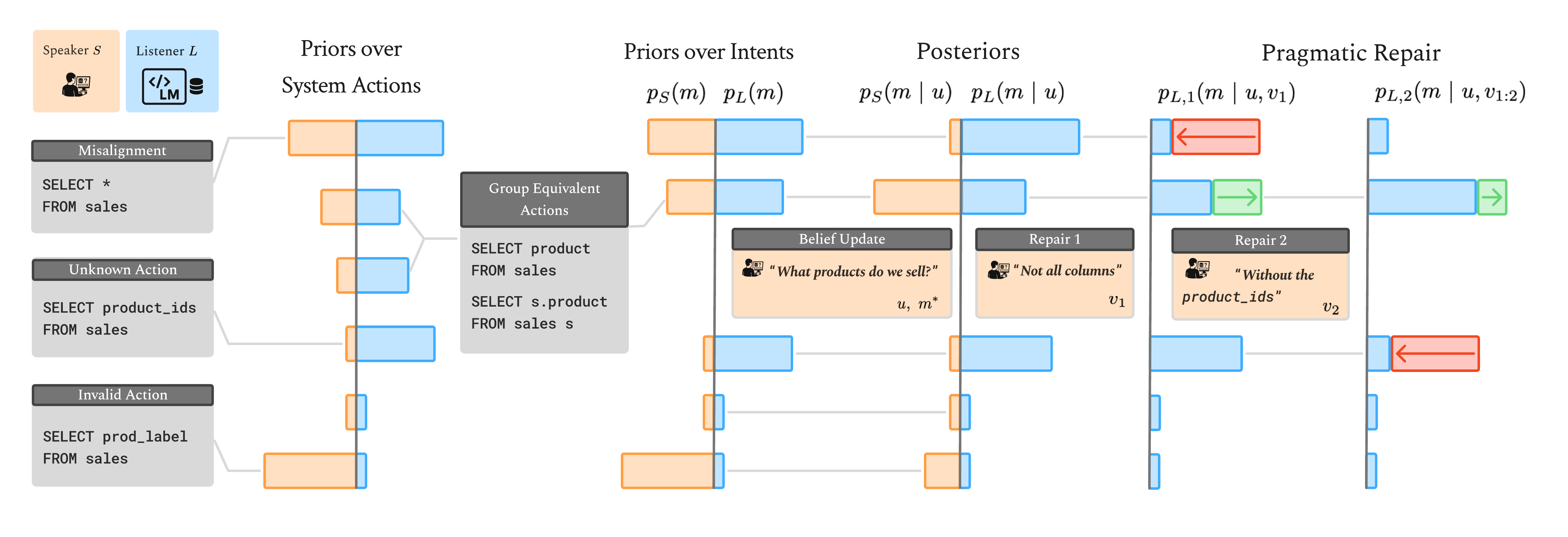}
    \vspace{-2em}
    \caption{Pragmatic repair in HCI, applied to the text-to-SQL task. A misalignment of priors over system actions, i.e., a user not knowing what actions are possible or considered probable by a system, amplifies ambiguity. }
    \Description{Diagram of pragmatic repair in text-to-SQL. 
    It shows how misalignment between user and system priors over actions and intents creates ambiguity. 
    Examples of system actions include valid, unknown, or invalid SQL queries. 
    Through belief updates and successive clarifications ("Not all columns", "Without the product\_ids"), 
    the user and system iteratively align their posteriors, narrowing toward the intended query.}
    \vspace{-1em}
    \label{fig:pragmatics}
\end{figure*}

\section{Pragmatic Repair for Natural Language Interfaces}
When users issue underspecified commands, natural language interfaces must somehow recover the intended meaning. We argue that this challenge can be understood through the lens of pragmatic inference: users economize their expressions, while systems operate on priors that may not align with user expectations. In this section, we formalize this misalignment and discuss pragmatic repair as a principled strategy for clarification, using the text-to-SQL domain as a running example for clarity.

\subsection{Pragmatic Inference in Natural Language Interfaces}

In natural language interfaces, a user provides a system with an utterance $u$, where their communicative intent $m^*$ is to receive their desired system outcome via a system action $a^* \in \mathcal A$.\footnote{The set $\mathcal A$ denotes \textit{executable system actions}.} For instance, in the text-to-SQL case shown in Figure~\ref{fig:pragmatics}, a user may ask a system $u=$``\textit{What products do we sell?}'' and expect a list of \verb|product|s from the \verb|sales| table as an outcome.
The system then infers a user intent $m$ given the user utterance and maps it to a suitable \emph{executable} system action $a \in \mathcal A$ that ideally achieves the user's expected outcome.
In the above example, this corresponds to finding the query $a^*=$ ``\verb|SELECT product FROM sales|''. We note that, generally, multiple system actions can fulfill the user intent if they are \textit{functionally equivalent}, such as the target query from above and ``\verb|SELECT s.product FROM sales s|'', in our example. 
Formally, the system induces a distribution 
\(
p(a \mid m), \; a \in \mathcal{A},
\)
over executable actions given an intent \(m \in \mathcal{M}\). 
The domain of this mapping is
\(
\mathrm{dom}(f) = \{\, m \in \mathcal{M} \mid \exists~a \in \mathcal{A}, \; p(a \mid m) > 0 \,\},
\)
i.e., the set of intents for which at least one executable action is supported by the system.
Thus, we can aggregate the prior probabilities of functionally equivalent programs into priors over intents $p_S(m)$ and $p_L(m)$.\looseness=-1

In natural language interfaces, a significant challenge is that the set of executable actions $\mathcal{A}$ is often not fully known to a user (cf. the \textit{gulf of execution} \cite{norman1986ucsd}).
This can result in a user expressing an intent $m^*$ that does not map to any executable system action, or formally, $m^* \notin \mathrm{dom}(f)$. In our example in Figure~\ref{fig:pragmatics}, the user may ask for the ``product label'', even though no ``\verb|prod_label|'' exists in the database.\looseness=-1

More generally, there may be misalignment between the user's and the system's priors over system actions. In other words, a user may be unaware of what system actions the system considers probable and may fail to estimate how well a produced utterance $u$ distinguishes $m^*$ from alternative probable intents. In the example of Figure~\ref{fig:pragmatics}, a system may be generally biased to show all columns, i.e., produce $a=$ ``\verb|SELECT * FROM sales|''. The user is unaware that this is something they need to distinguish against and assumes that their utterance $u=$``\textit{What products do we sell?}'' yields a peaked posterior $p_L(m~|~u)$ at $m^*$. However, to the system, the utterance is ambiguous; its resulting posterior $p_L (m\mid u)$ is high-entropy, and sampling from it leads to unexpected behavior.

Thus, communicative efficiency in natural language interfaces depends not only on the informativeness of utterances under shared priors, but also on aligning the priors themselves. 
This alignment can be facilitated by (a) surfacing the action space $\mathcal{A}$ to the user, thereby reducing uncertainty about what outcomes are possible, and (b) adapting or calibrating the system’s prior to better reflect the user’s expectations, for instance, through clarification mechanisms.

\begin{figure*}[t!]
    \centering
    \includegraphics[width=\linewidth]{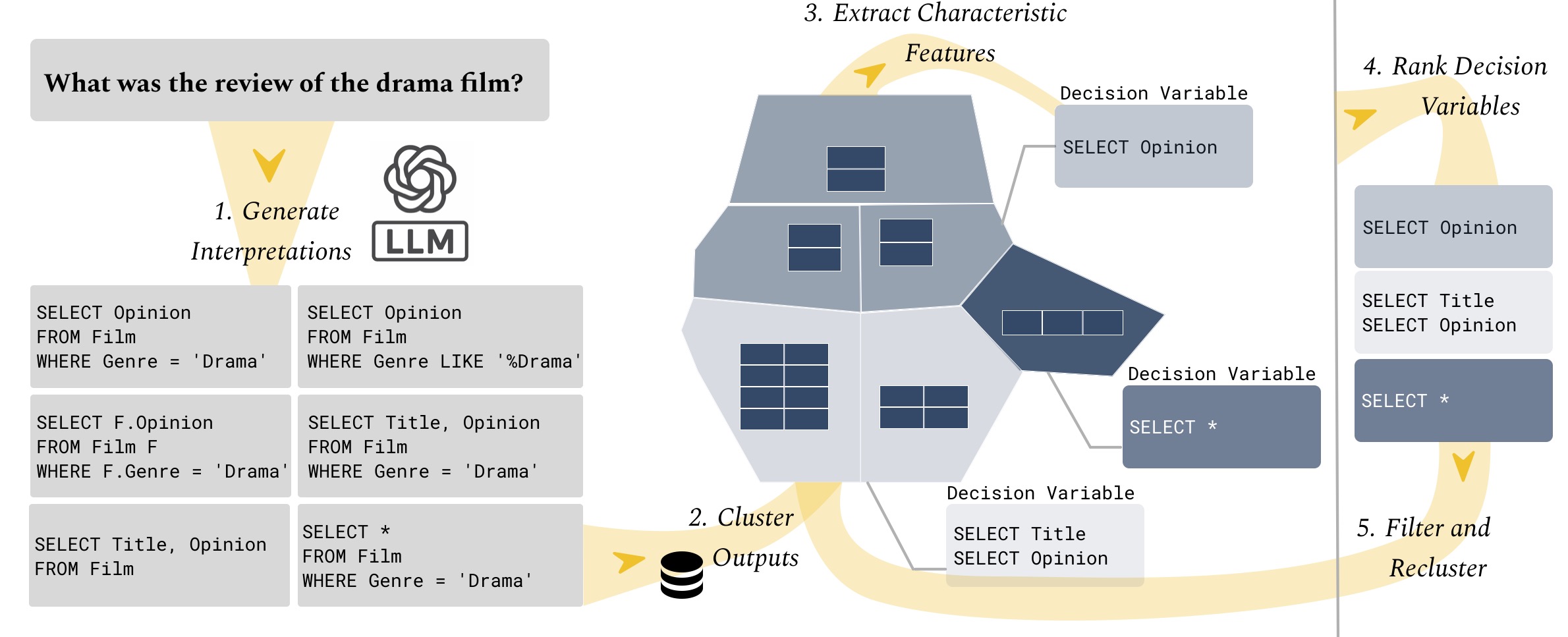}
    \vspace{-1em}
    \caption{Visualization of the algorithm described in Section~\ref{sec:algorithm} applied to text-to-SQL. First, an LLM generates diverse outputs, which are then embedded and clustered together. For each cluster, we extract the most characteristic feature based on the lift. Finally, we order decision variables based on expected information gain. Once a user has made a decision, the candidate set is filtered, and the remaining candidates are reclustered. }
    \Description{Visualization of the PleaSQLarify algorithm. 
    A user asks a question, and the LLM generates multiple SQL interpretations. 
    These outputs are clustered, and characteristic features are extracted to identify decision variables 
    such as SELECT Opinion, SELECT Title, or SELECT *. 
    Decision variables are ranked by expected information gain, and users iteratively filter and recluster 
    the candidate set until the intended query is isolated.}
    \vspace{-1.5em}
    \label{fig:algorithm}
\end{figure*}

\subsection{Interactive Utterance Grounding as Pragmatic Repair} \label{subsec:repair}
While surfacing the available action space $\mathcal{A}$ can help users form more accurate queries, directly exposing the full set of actions is often impractical and cognitively costly, especially when $\mathcal{A}$ is large or highly technical.
Naturally, the question arises whether---akin to utterance grounding in pragmatics---underspecified user queries can also be interactively grounded in meaning through such clarifying repairs.
To this end, a system may leverage the posterior distribution $p_L(m \mid u)$ from the user’s initial utterance to focus only on actionable candidate meanings. This reframes the problem from ``searching $\mathcal{A}$'' to ``disambiguating among a small set of likely intents,'' allowing ambiguity to be resolved through targeted, incremental clarification.
Formally, as shown in Figure~\ref{fig:pragmatics}, let $\mathcal M_0 = \{m : p_L(m \mid u) > 0\}$
be the set of \textit{candidate} intents after the initial utterance $u$. At each interaction turn $t$, the system chooses a decision variable
$Z_t : \mathcal M_{t-1} \rightarrow \mathcal{V}_t$
---a salient dimension along which the candidates differ—and presents the user with alternatives $v \in \mathcal{V}_t$.
Crucially, to minimize the number of interaction turns, the system selects the variable $Z_t$ that maximizes the expected information gain with respect to the intent distribution.
The user’s choice $v_t \in \mathcal{V}t$ yields the updated candidate set
$\mathcal M_t = \{m \in \mathcal M_{t-1} \mid Z_t(m) = v_t\}$.
This refinement corresponds to incrementally conditioning the posterior on additional evidence:
\begin{equation}
p_{L,t}(m \mid u, v_{1:t}) \propto p_L(m)\,p_S(u \mid m)\,\prod_{i=1}^t \mathbbm{1}\{Z_i(m) = v_i\}.
\end{equation}
Over successive turns, the candidate set narrows through targeted, pragmatically relevant repair moves, mirroring how human speakers iteratively refine an underspecified referring expression until both parties can treat the meaning as mutually accepted.

\subsection{Design Implications for Pragmatic Repair}\label{subsec:design-implications} 
The prior exposition informs us about the design requirements for a system that allows for pragmatic repair. 

\begin{enumerate}
    \item[\textbf{R1}] Overall, the system should surface the available action space $\mathcal A$ in an incremental, human-understandable way.
    \item[\textbf{R2}] Assuming the user's goal is to achieve a specific outcome $m^*$, clarifications should distinguish between \emph{intents} rather than system actions, since differentiating between functionally equivalent actions is irrelevant to the user’s goal.
    \item[\textbf{R3}] The pragmatic principle of least collaborative effort motivates minimizing the number of clarifying interactions by prioritizing asking the user \emph{informative} clarifications, i.e., choosing clarifications which maximally reduce the uncertainty about the intended meaning.
\end{enumerate}

\section{An Algorithm for Pragmatic Repair in Natural Language Interfaces}
\label{sec:algorithm}

In the following, we describe an algorithm that implements pragmatic repair for natural language interfaces. Following the design principles specified in Section~\ref{subsec:design-implications}, we represent the action space by (1) generating a set of probable system actions, (2) aggregating and clustering functionally similar groups, (3) extracting decision variables based on the most characteristic features in these clusters, and (4) extracting and surfacing the most informative features via information gain.\looseness=-1

\paragraph{1. Generating a Set of Probable Actions}
To support the wide exploration of system interpretations of the input utterance $u$, we first generate a finite candidate set of $N$ probable system actions $\mathcal{A} = \{a_1, \ldots, a_N\}$ under the input utterance.
The mechanism for generating $\mathcal{A}$ is domain-dependent, and for language model-based tasks, it may be attained by resampling a language model conditioned on the input utterance $u$ that contains the task instruction, i.e., $\mathcal A = \{a_i \sim p_{\mathrm{LM}}(a \mid u)\}_{i=1}^N$.

\paragraph{2. Functional Clustering}
As discussed in Section~\ref{subsec:pragmatics}, multiple different system actions can realize the same intent $m$, i.e., execute the same function.
We formalize this by saying that two system actions are \textit{functionally equivalent} if they produce the same functional outcome. 
Let $m(a, \mathcal C)$ denote the outcome of executing action $a$ in context $\mathcal C$. Two actions $a_i$ and $a_j$ are functionally equivalent with respect to $\mathcal C$ if $m(a_i, \mathcal C) = m(a_j, \mathcal C)$.
We further introduce the \emph{functional similarity} $S(a_i, a_j)$ between two system actions as a similarity score between $m(a_i, \mathcal C)$ and $m(a_j, \mathcal C)$. If $S(a_i, a_j) \approx 1$, the results are nearly identical.
Using this similarity measure, we cluster candidate actions to produce semantically coherent regions of the search space. 

\paragraph{3a. Extracting Atomic Decision Variables} 
To discriminate between candidate actions in $\mathcal A$, we represent each action as a $d$-dimensional binary \emph{atomic feature vector} $\mathbf{z}(a) \in \{0, 1\}^d,$
where each dimension corresponds to the presence of a specific atomic action component. 
These atoms are the smallest distinguishable units between actions and enable incremental, interpretable decision-making (\textbf{R1}). 
Later, these atomic features will be combined into higher-level, human-interpretable decision variables $Z_t$ as described in Section~\ref{subsec:repair}.

\begin{figure*}[t!]
    \centering
    \includegraphics[width=\linewidth]{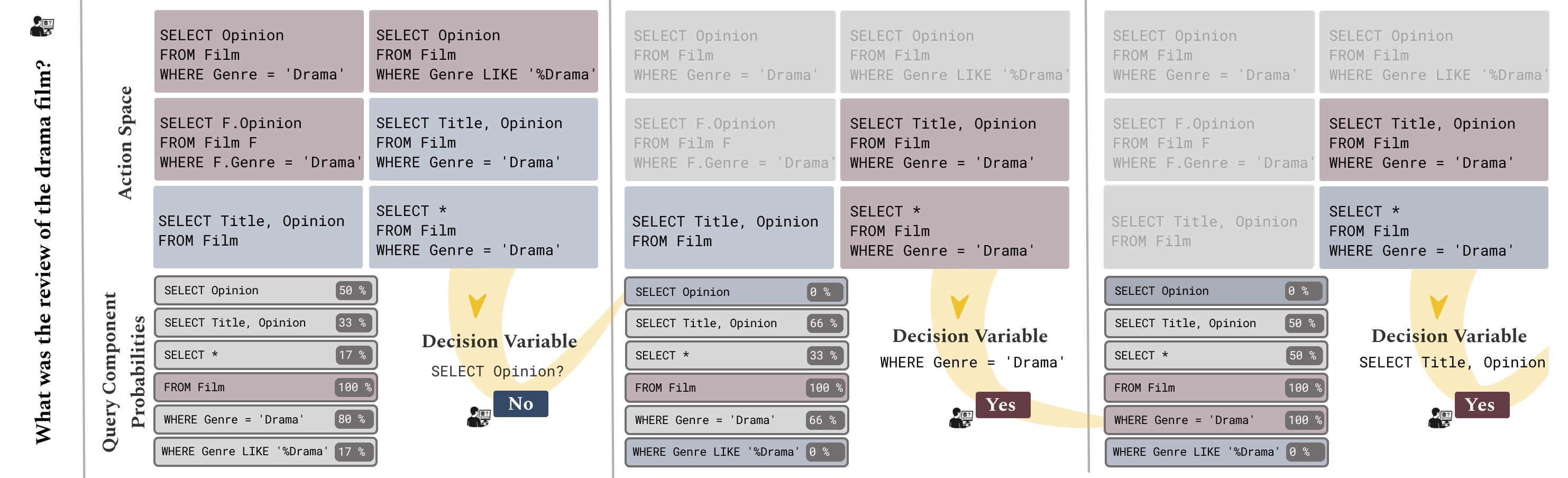}
    \vspace{-1em}
    \caption{Iterative candidate filtering and query population using the proposed algorithm  Section~\ref{sec:algorithm} applied to text-to-SQL. 
    The initial candidate set is iteratively filtered through targeted user feedback on the decision variables proposed by our algorithm until the candidate set consists of a single query, i.e., until the user has identified the query that most closely matches their intent.
    }
    \Description{Visualization of the PleaSQLarify clarification process.
Given a user query, the system generates multiple SQL candidates.
The user responds to targeted clarification prompts based on decision variables that partition the candidate space, progressively narrowing it until only one executable query remains.}
    \vspace{-1.5em}
    \label{fig:filtering}
\end{figure*}

\paragraph{3b. Grouping Atomic Decision Variables} 
While static, atomic features are useful internal discriminators between system actions, they have two limitations:  
(1) \emph{Equivalence masking}---vastly different atomic feature sets may yield functionally similar outputs;
(2) \emph{Interaction neglect}---meaningful differences can emerge only when features are considered jointly, i.e., their semantic meaning emerges compositionally.
To address this, we aim to construct grouped decision variables that are both useful \textit{functional} discriminators, while remaining \textit{semantically interpretable} to a user and, thus, exposing the available action space in a meaningful way (\textbf{R1}, \textbf{R2}).
Building on top of the clusters obtained in step 2, we extract the most characteristic features for each cluster.
For a group of atomic features $g$ within a cluster $C$, define
\begin{equation}
p_{\mathrm{in}}(g) \;=\; \frac{\#\{\;a \in C \mid g \in \mathbf{z}(a) \;\}}{|C|},
\quad
p_{\mathrm{all}}(g) \;=\; \frac{\#\{\;a \in \mathcal A \mid g \in \mathbf{z}(a) \;\}}{|\mathcal A|},
\end{equation}
where $\mathcal A$ is the (finite) set of probable system actions.  
We then compute the \emph{lift}:
\begin{equation}
\mathrm{lift}(g, C) \;=\; \frac{p_{\mathrm{in}}(g)}{p_{\mathrm{all}}(g)}.
\end{equation}
A lift greater than $1$ means $g$ is more prevalent \emph{within} the cluster than expected from its global frequency, thus making it a \textit{characteristic} set of features.
Finally, to capture implicit dependencies between features, we compute the empirical co-occurrence
\begin{equation}
p\big(z_{Z} = 1 \,\big|\, z_{g} = 1\big) 
\;=\;
\frac{\#\{\, a \in \mathcal A \mid z_{g}(a) = 1 \ \text{and} \ z_{Z}(a) = 1 \,\}}{\#\{\, a \in \mathcal A \mid z_{g}(a) = 1 \,\}},
\end{equation}
where $z_{g}(a)$ is the binary indicator that query $a$ contains feature group $g$, and $z_{Z}(a)$ is the binary indicator that it contains feature $Z$. 
This quantity represents the probability that $Z$ is present given that $g$ appears, thus identifying \emph{implicitly included} variables.

\paragraph{4. Ranking Decision Variables} 
Let $\mathcal{M}_t$ be the set of intents that are still consistent with the user’s clarifications at step $t$, and let $p_t(m)$ be the current belief distribution over $\mathcal{M}_t$. Following requirement \textbf{R3}, the clarifications should be sought that maximally reduce the expected semantic uncertainty in the posterior over meanings. This is exactly the principle behind information gain \cite{mitchell1997machine}.
Namely, for each candidate decision variable $Z$, the \emph{expected information gain} at time $t$ is:
\begin{equation}
\mathrm{IG}_t(Z) \;=\; H(p_t) \;-\; \sum_{v \in \mathcal{V}(Z)} P_t(Z = v) \, H\!\left(p_t(\,\cdot\, \mid Z = v)\right),    
\end{equation}
where $H(p_t) \;=\; -\sum_{m \in \mathcal M_t} p_t(m) \,\log p_t(m)$ is the Shannon entropy and $ P_t(Z = v) \;=\; \sum_{m \in M_t} p_t(m) \,\mathbbm{1}\!\{ Z(m) = v \}.$
Using the expected information gain as an optimization criterion, we select the next decision variable for clarification as:\looseness=-1
\begin{equation}    
Z_t^\ast \;=\; \arg\max_Z \; \mathrm{IG}_t(Z).
\end{equation}

\paragraph{5. Filtering and Reclustering}
After the user provides feedback on decision variable $Z_t^\ast$, we filter $\mathcal{A}_t$ to retain only actions consistent with the stated preference and update the belief distribution $p_t$ accordingly. The reduced candidate set $\mathcal A_{t+1} \subset \mathcal A_t$ is reclustered, and the algorithm restarts at step 2. This loop continues until a unique action remains in the candidate set.

\begin{figure*}[th!]
    \centering
    \includegraphics[width=\linewidth]{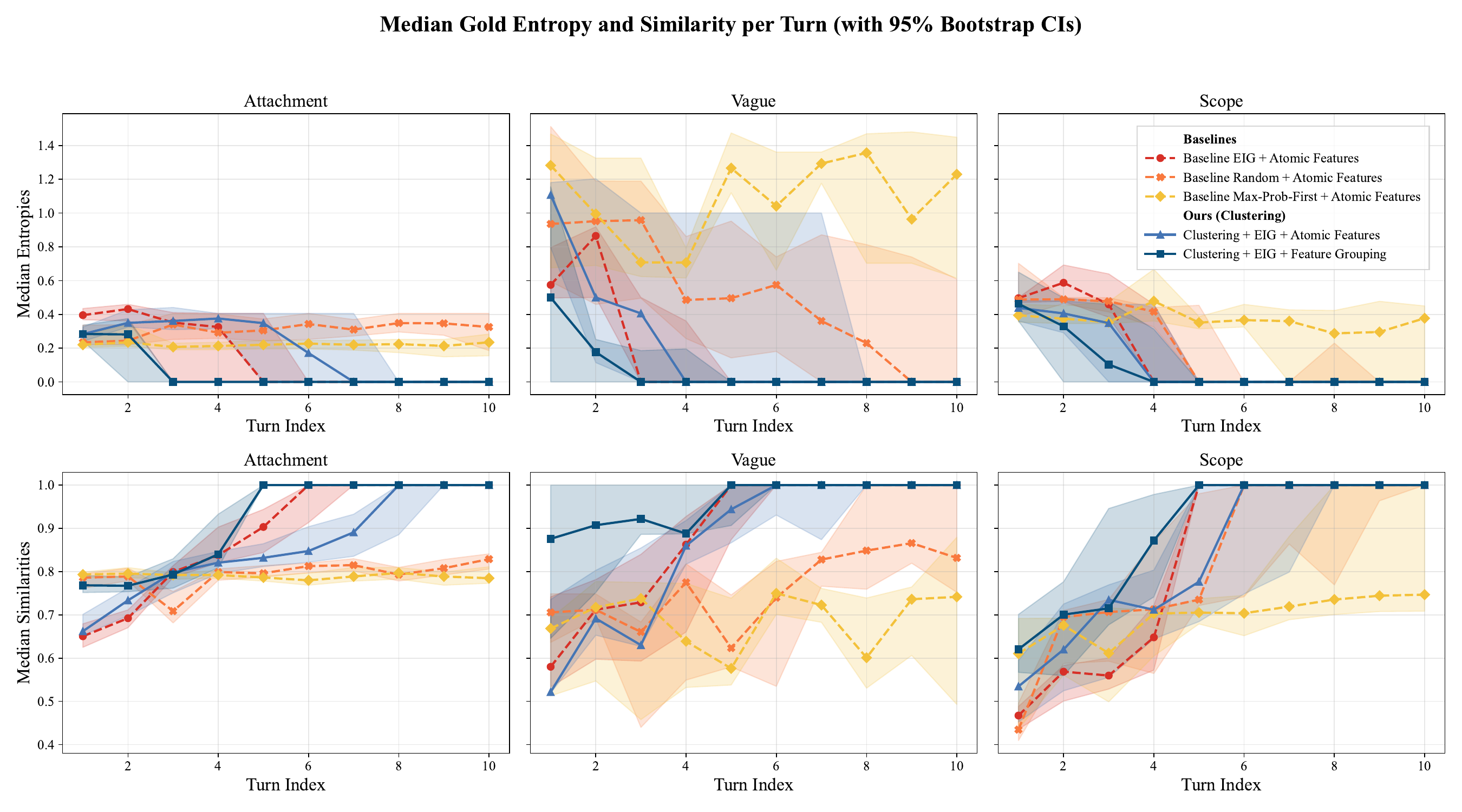}
    \vspace{-2em}
    \caption{
        Median per-turn gold-label entropy (top) and functional output similarity (bottom) across three ambiguity types in the AMBROSIA dataset. Clustering-based methods (solid blue) resolve uncertainty more rapidly and converge to functionally coherent hypotheses in fewer turns than baselines (dashed red/orange).
    }
    \Description{Six line plots compare baselines and clustering-based methods across three ambiguity types 
    (Attachment, Vague, Scope). The top row shows median entropy decreasing over turns, while the bottom row 
    shows median similarity increasing. Clustering-based methods (solid blue) reduce uncertainty faster and 
    achieve higher similarity than baselines (dashed red/orange). Shaded regions indicate 95\% confidence intervals.}
    \vspace{-1em}
    \label{fig:entropies_sims}
\end{figure*}

\section{Algorithm Application}

\paragraph{An Instantiation for Text-to-SQL}
For the remainder of the paper, we instantiate the pragmatic repair algorithm for text-to-SQL, where users pose natural language questions to be translated into executable database queries. We visualize each step of the algorithm in Figure~\ref{fig:algorithm} and the higher-level iterative clarification process in Figure~\ref{fig:filtering}.
In this domain, we generate the action space by sampling different SQL interpretations from a large language model under the input utterance (cf. step 1, Figure~\ref{fig:algorithm}). 
To cluster system actions functionally, we use \textit{execution similarity} as a similarity criterion; we consider two SQL queries functionally similar if they produce a similar output when run against some test database $\mathcal D$ (cf. step 2, Figure~\ref{fig:algorithm}).
Note that functional equivalence in this setting is \emph{approximate}: two SQL queries are strictly equivalent iff they yield identical outputs for all possible databases~\cite{köberlein2025qss}.
To construct decision variables, we parse each query into an abstract syntax tree and encode clause elements (e.g., \texttt{GROUP BY column\_x} or \texttt{SELECT column\_y}) as binary features. 
The remainder of the algorithm is agnostic to the application domain; for a given binary atomic feature vector, we group, extract, and rank characteristic features exactly as outlined in  Section~\ref{sec:algorithm} (cf. steps 3-5 in Figure~\ref{fig:algorithm}).

\paragraph{Transferability to other Domains}
As showcased above, extracting relevant features and clustering them by functional similarity requires task-specific adaptation, whereas the subsequent ranking and selection of decision variables operate task-agnostically on the generated atomic feature vector.
As such, the algorithm applies more broadly to natural language tasks where system outputs can be decomposed into \textit{comparable atomic components} that are informative of some notion of \textit{functional interpretation}.
This decomposition is straightforward when outputs follow a compositional grammar with a simple functional evaluation, as in SQL, but it becomes challenging for unstructured or free-form generation.
In the latter case, prior work has demonstrated the efficacy of task-specific taxonomies that specify and measure components contributing to diverging input interpretations.
For instance, in open-domain question answering \cite{ma2025ambigchat, kim2023tree} the input ambiguity may be systematically decomposed into discrete facets along which interpretations diverge---such as entity, temporal, and geographical specificity---enabling hierarchical disambiguation even without formal syntax.
Similarly, tools that handle ambiguity for natural language data visualization \cite{setlur2019inferencing, gao2015datatone, inan2025identifying} use such taxonomies to identify axes of data and visual underspecification that may lead to unintended outcomes.

Beyond extracting meaningful features, our approach relies on the extracted atomic components being interpretable to the target user, as users are asked to provide clarification directly on the extracted decision variables. 
In our text-to-SQL instantiation, we surface SQL clauses as decision variables, making clarification accessible primarily to SQL-literate users.
However, strategies such as representing the decision variable through a natural language question \cite{liu2023abstraction, narechania2021diy, ma2025ambigchat} have been shown to help lay users understand even more complex distinctions.

\begin{figure*}[th!]
    \centering
    \includegraphics[width=\linewidth]{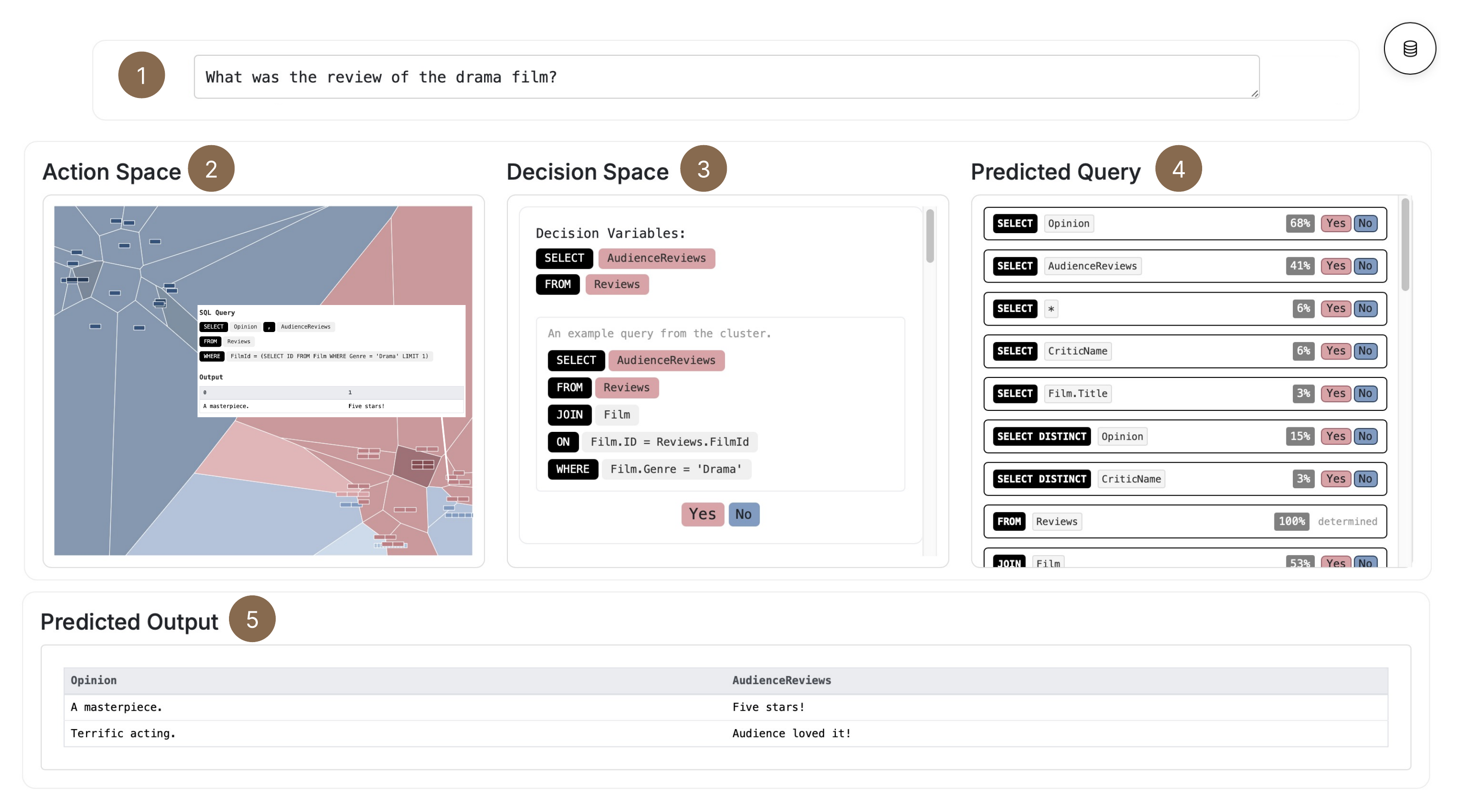}
    \vspace{-2em}
    \caption{
        The visual interactive interface consists of the user's input/utterance field (1), three main analysis views, i.e., the \textit{Action Space} (2) for exploration, \textit{Decision Space} (3) for decision making, and \textit{Predicted Query} (4) for confirmation tasks, and the \textit{Predicted Output} view (5) showing the predicted result of the database. 
    }
    \Description{The interactive interface for PleaSQLarify. 
    It includes (1) a user input field for natural language interfaces, 
    (2) an Action Space visualization showing diverse SQL candidates, 
    (3) a Decision Space view where users select among decision variables, 
    (4) a Predicted Query panel with confidence scores and Yes/No confirmations, 
    and (5) a Predicted Output table showing the retrieved database results.}
    \vspace{-1.2em}
    \label{fig:interface}
\end{figure*}

\section{Quantitative Evaluation}\label{sec:evaluation}

We first provide a brief quantitative evaluation of the algorithm outlined in  Section~\ref{sec:algorithm} for text-to-SQL to highlight its ability to discriminate between semantically different programs (\textbf{R2}) within a minimal number of user interactions (\textbf{R3}).\looseness=-1

\paragraph{Setup}

We evaluate our algorithm on the AMBROSIA dataset \cite{saparina2024ambrosia}---a text-to-SQL dataset designed to contain various types of ambiguity. Namely, they differentiate between \textit{scope} and \textit{attachment} ambiguity, as well as \textit{vagueness}. For each sample, we generate a set of candidate queries $\mathcal A$ by prompting \verb|GPT-4o| \cite{openai2024gpt4technicalreport} $N = 50$ times\footnote{We drop syntactically invalid candidate queries from the candidate set, as these are not part of the action space $\mathcal A$. The effective number of considered candidates is, therefore, $ \leq 50$.} with the ambiguous query at high temperature\footnote{We used 0.7 to balance the syntactical correctness of the generated SQL with the diversity of generated programs.}.
We extract the abstract syntax tree from each generated query using the spider SQL parser \cite{yu2018spider}. Each value in the AST is then encoded as a binary feature, which results in a binary feature matrix $\{\mathbf{z}(a)\}^N$.
We construct the output similarity matrix $S$ by executing each query against the test database provided by AMBROSIA and computing embedding-based output similarity using \verb|all-MiniLM-L6-v2| \cite{reimers2019sentencebert}. We cluster outputs using hierarchical clustering \cite{kaufman1990finding}, as this is a deterministic algorithm that allows us to exactly specify the number of clusters.
As the dataset provides two (or more) gold interpretations per sample, we can additionally manually assign each generated candidate query to either of the high-level gold intents. As disambiguation baselines, we implement (1) random decision variable choice, (2) a greedy decision variable selection, i.e., choosing the decision variable that splits on the value with the highest current posterior probability first, and (3) expected information gain without functional clustering. We compare this to two variants of our clustering-based algorithm (atomic vs. grouped decision variables) by selecting one of the gold labels as the user intent and choosing the decision variable value that agrees with their intent. The interaction algorithm terminates once all remaining programs are functionally equivalent.

\paragraph{Results}

Figure~\ref{fig:entropies_sims} shows how different clarification strategies affect the pace of disambiguation over successive turns in interactive text-to-SQL repair.
The top row reports the median gold-label entropy over the candidate meaning distribution, a direct measure of residual semantic uncertainty.
The bottom row tracks the median functional similarity within the remaining candidate set, reflecting how tightly the surviving queries cluster in execution space.

Across all three ambiguity types, the two proposed clustering-based strategies consistently outperform the baselines.
They achieve steeper entropy declines, reaching low uncertainty levels in fewer turns, and attain higher within-branch functional similarity earlier in the interaction.
This behavior aligns with our design requirement~\textbf{R3}: prioritizing clarifications that maximally reduce semantic uncertainty, thereby minimizing the number of required repair turns.
In particular, running expected information gain on grouped features rapidly collapses the hypothesis space—often to a functionally homogeneous set—within three to five turns, whereas baseline methods can require many more interactions to achieve comparable certainty.
The results suggest that surfacing semantically meaningful, high-information decision variables not only accelerates the grounding of underspecified utterances but also yields candidate sets that are more coherent in functional terms, enabling quicker and more confident resolution of user intent.

\begin{figure*}[th!]
    \centering
    \includegraphics[width=0.945\linewidth]{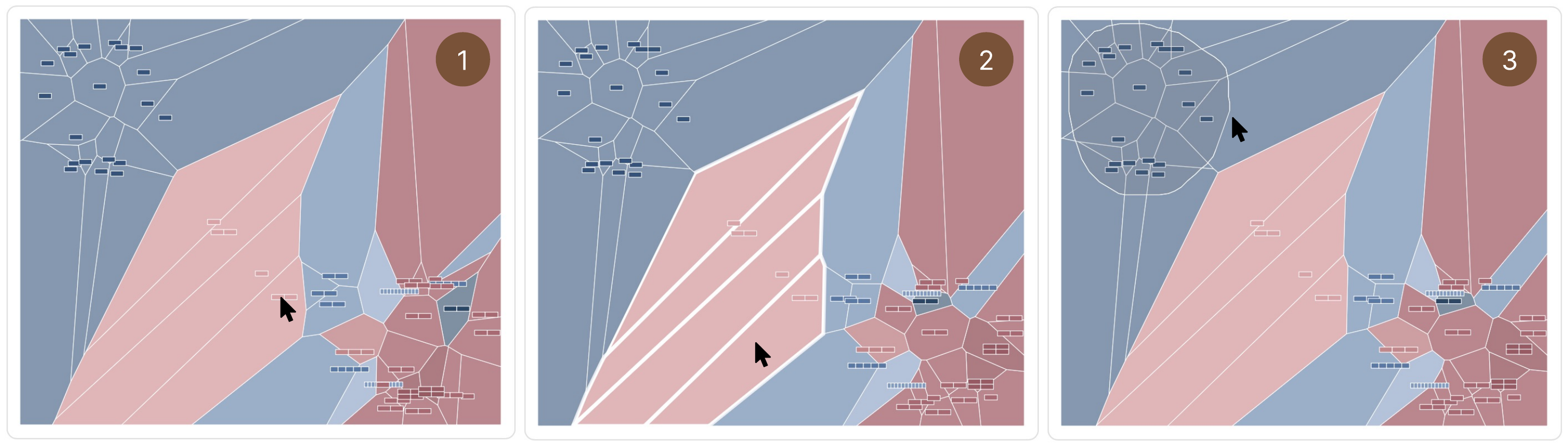}
    \vspace{-1em}
    \caption{
        The \textit{Action Space} is a UMAP projection of the SQL queries visualized as glyphs and surrounded by Voronoi cells. Their colors correspond to the cluster indices, whereby red and blue shades represent the queries that include the decision variables and those that exclude them. The main purpose of this view is to support data exploration. The users can filter interesting queries by either clicking on a query glyph (1), a cluster of query glyphs (2), or by selecting a group of query glyphs with a lasso (3). 
    }
    \Description{The Action Space visualized as a UMAP projection of SQL queries, shown as glyphs within Voronoi cells. 
    Queries are colored red or blue depending on whether they include specific decision variables. 
    Users can explore by selecting a single query glyph, a cluster of queries, or a group of queries with a lasso tool.}
    \vspace{-1em}
    \label{fig:interactions-action-space}
\end{figure*}

\section{Visual Interface}\label{sec:design}

While atomic decision variables can be predictive or informative, they often correspond to latent factors that are not readily interpretable. Moreover, exposing decision variables alone does not provide the user with a clear view of the available action space $\mathcal{A}$.
To address these remaining challenges, we build a visual interface that aims to (1) surface the system's probable action space $\mathcal A$ by visualizing a semantically clustered representation of the remaining candidate set $M_t$ at each decision step $t$ (\textbf{R1}), (2) make user decisions more interpretable by surfacing grouped decision variables and implicitly included variables (\textbf{R2}), as derived in  Section~\ref{sec:algorithm}, and (3) support efficient intent clarification with minimal iterations (\textbf{R3}), while recording decisions and their effects on the final query.

We note that the interface is primarily built as a research tool designed to understand users' workflows during utterance disambiguation in text-to-SQL.
While not immediately intended for general LLM end-users, we could imagine that individual interface components or our insights may inform future designs for utterance disambiguation in natural language interfaces.
We further address this in our limitations in  Section~\ref{sec:limitations}.

\subsection{Design Rationale}\label{sec:design-rationale} 

As shown in~Figure~\ref{fig:interface}, our visual interactive interface is structured into three parts: the user’s utterance is positioned at the top of the page, the central region presents the three query refinement components, and the bottom of the page displays the intermediate predicted results table. The three core components for query refinement, i.e., the \textit{Action Space}, \textit{Decision Space}, and \textit{Predicted Query}, address the system requirements introduced in Section~\ref{subsec:design-implications}.
The primary objective of the \textit{Action Space} is to assist users in exploring alternative interpretations and manually filtering queries to those that are most relevant.
The \textit{Decision Space} assists users by suggesting decision variables that can disambiguate utterance meanings in the most efficient manner.
Finally, the \textit{Predicted Query} supports the provenance of the decisions made. In addition, this view provides an overview of the most probable atomic features in the final query, conditioned on the user's decisions.
Because the three views are interlinked, a user’s interaction in one view triggers corresponding updates in the others.
In particular, the query filtering in the \textit{Action Space} shapes the set of subsequent suggested decision variables.
By accepting or rejecting a decision variable, queries that include or exclude particular features will be filtered, and the corresponding views updated.
In the following, we describe design considerations to support the usability of these core components of the interface.

\paragraph{Query Visualization} 
SQL queries are the main building blocks of the interface.
In \textsc{PleaSQLarify}, a query has two \setlength{\columnsep}{5pt} types of representation. 
As shown in~Figure~\ref{fig:teaser}, by default, the query is visualized
as a list of atomic features, where a feature is built from a keyword (a span with a black background, e.g., \verb|SELECT|) and a value (a span with a light gray background, e.g., \verb|Opinion|). 
In addition to this visualization of atomic features, we use a more compact representation to give an overview of the probable action space $\mathcal A$. 
In this compact form, the aim is to highlight the differences in
outputs produced by different queries. 
Thus, the query is represented as a glyph, which mimics its output table retrieved from the
database (number of rows and columns, scaled to a maximum size of 50 px)~
\aptLtoX{\raisebox{-3pt}{\includegraphics[width=0.022\textwidth]{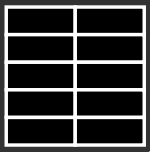}}}{\inlinegraphics{figures/glyph-black.png}}.
The glyphs are intended to emphasize only the presence of output differences; thus, users are not expected to examine precise row and column counts, which would be difficult for larger~tables.

\begin{figure*}[t]
    \centering
    \includegraphics[width=0.996\linewidth]{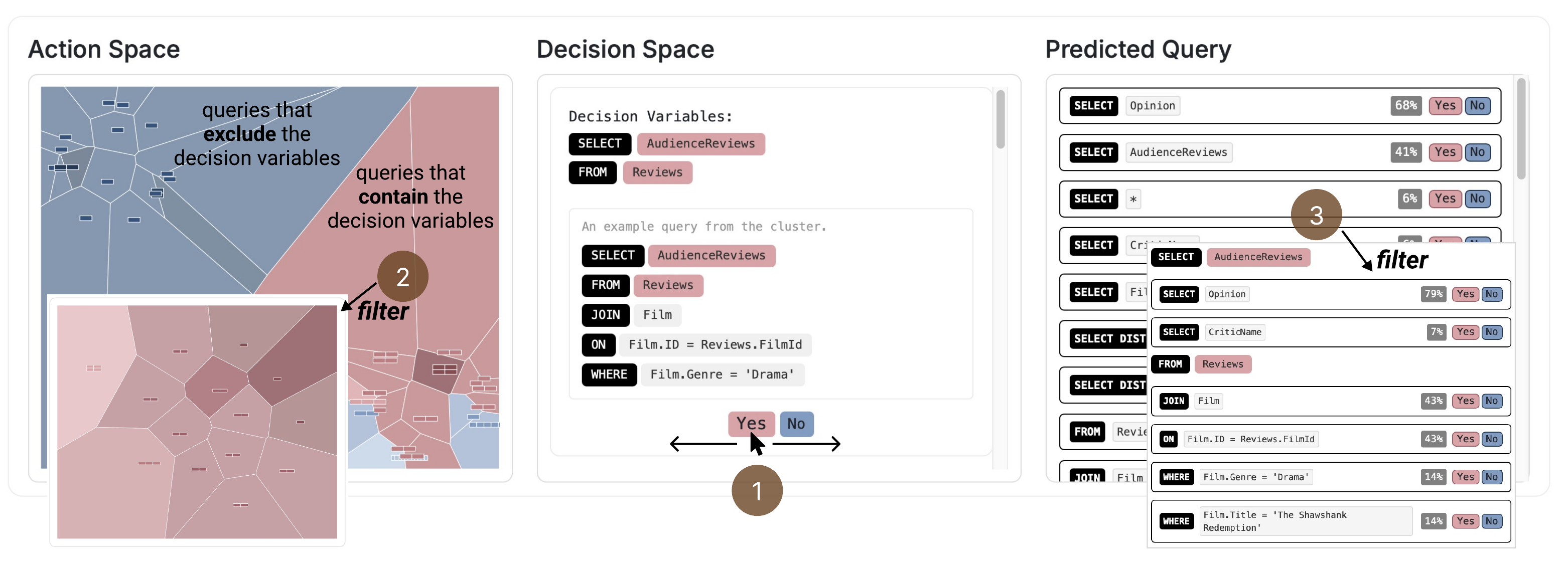}
    \vspace{-2em}
    \caption{When the user accepts a decision variable by clicking on the ``Yes'' button in the \textit{Decision Space} (1), the queries that contain/exclude the variables will be filtered in the \textit{Action Space} (2).
        In addition, the \textit{Predicted Query} panel will be updated with atomic features that are likely to occur in the final query (3).}
    \Description{Interface view showing how selecting a decision variable updates the query space. 
    When the user clicks “Yes” in the Decision Space, the Action Space is filtered to highlight queries 
    that include the chosen variable, and the Predicted Query panel is updated with corresponding features 
    and likelihoods for the final query.}
    \vspace{-1.2em}
    \label{fig:decision}
\end{figure*}

\paragraph{Global Color Encoding} 
The primary purpose of the interface is to clearly convey the decision variables to users and illustrate how accepting or rejecting them influences the next possible choices.
We thus use a global binary color encoding to make the impact of the decisions apparent. 
We use a red shade instead of light gray to represent the atomic features that act as current decision variables~\aptLtoX{\raisebox{-3pt}{\includegraphics[width=0.2\textwidth]{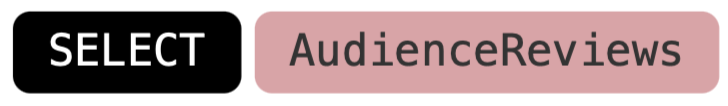}}}{\inlinegraphics{figures/decision_variable.png}}. In addition, we use a red color scale~\aptLtoX{\raisebox{-1pt}{\includegraphics[width=0.06\textwidth]{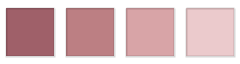}}}{\inlinegraphicsSmall{figures/red.png}} to color queries (i.e., query glyphs~\aptLtoX{\raisebox{-3pt}{\includegraphics[width=0.022\textwidth]{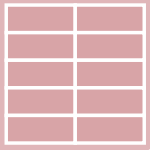}}}{\inlinegraphics{figures/glyph.png}}) that would be filtered if the user accepts the suggested decision variable, whereby the different shades of red represent different clusters determined by our algorithm (see~ Section~\ref{sec:algorithm}).
Red was selected because previous research indicates that this color functions as a signal of relevance, conveying that a stimulus is important and merits attention~\cite{buechner2014red}.
A blue color scale~\aptLtoX{\raisebox{-1pt}{\includegraphics[width=0.06\textwidth]{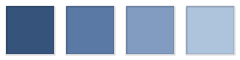}}}{\inlinegraphicsSmall{figures/blue.png}} is used to color queries (i.e., query glyphs~\aptLtoX{\raisebox{-3pt}{\includegraphics[width=0.022\textwidth]{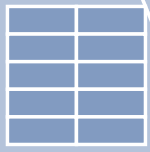}}}{\inlinegraphics{figures/glyph-blue.png}}) that exclude the current decision variable. 

\subsection{Visual, Interactive Components}
We now discuss the visual and interaction approaches used in the three core views of the interface: the \textit{Action Space}, the \textit{Decision Space}, and the \textit{Predicted Query}. 

\paragraph{Action Space} 
The purpose of the \textit{Action Space} is to give an overview of all possible system actions, i.e., LLM-generated queries for the given user's utterance, and to enable users to selectively filter queries for relevance.
As described in~ Section~\ref{sec:design-rationale}, each individual query is represented through a glyph that mimics its output from the database.
The two-dimensional positions of the glyphs are determined by applying the UMAP projection algorithm~\cite{mcinnes2018umap} to the distance matrix of queries, the calculation of which is explained in~ Section~\ref{sec:algorithm}.
As mentioned in~ Section~\ref{sec:design-rationale}, the query glyphs are colored red if they contain the current decision variable displayed in the \textit{Decision Space} (see the next paragraph) and blue otherwise. 
Moreover, we use different shades to highlight the clusters of queries with similar database results.
We enhance cluster readability by enclosing each glyph within a Voronoi cell, which is a common design choice in visual interactive labeling interfaces to illustrate class boundaries~\cite{chen2025activiz}, and applying the same color to the surrounding polygon as the query glyph~\cite{Aurenhammer1991voronoi}. 

We use multiple interaction methods to enable query exploration.
The user can hover over a query glyph to display a tooltip showing the query's atomic features and the output table (see an example in~Figure~\ref{fig:interface}).
As the user explores the \textit{Action Space} and identifies queries that likely reflect their intended meaning, we enable them to narrow the space down to the most relevant queries.
As shown in~Figure~\ref{fig:interactions-action-space}, the queries can be filtered by applying three different interaction methods.
If the query matches the user's interpretation, it can be selected by clicking on the particular glyph. 
By hovering over Voronoi cells, the corresponding cluster will be highlighted through white polygon borders.
To select a cluster, the \textit{shift} key needs to be pressed while clicking on one of the Voronoi cells in the cluster. Finally, the user can also use the lasso functionality to select a desired group of query glyphs.

\begin{figure*}[th!]
    \centering
    \includegraphics[width=\linewidth]{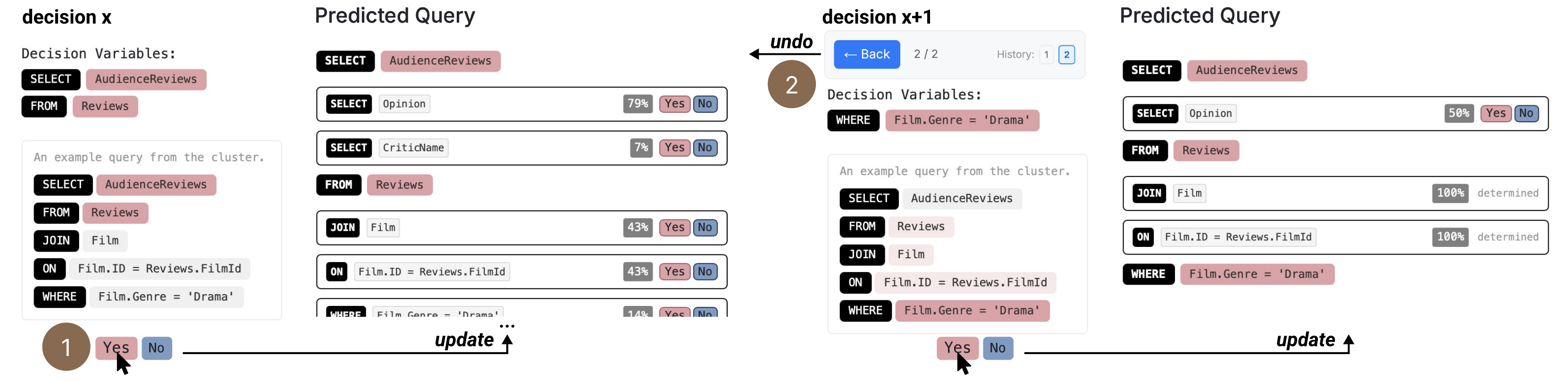}
    \vspace{-2em}
    \caption{
        Every time a decision is made by the user (1), the number of atomic features in the \textit{Predicted Query} is reduced. In particular, the features are retained only if they are included in the filtered queries. We color the selected decision variables and update the probabilities for the remaining ones. In case a feature has a 100\% probability to be part of the final query, it is marked as \textit{determined}. Every decision made can be reversed by clicking on the \textit{Back} button displayed in the \textit{Decision Space} (2).
    }
    \Description{Example of iterative decision-making in the interface. 
    At each step, the user selects a decision variable, and the Predicted Query panel is updated. 
    Atomic features not included in the filtered queries are removed, probabilities for remaining features are adjusted, 
    and features with 100\% probability are marked as determined. Every decision made can be reversed by clicking on the Back button displayed in the Decision Space.}
    \vspace{-1.2em}
    \label{fig:decision-predicted-query}
\end{figure*}

\paragraph{Decision Space} 
The goal of the \textit{Decision Space} is to guide users through decisions that allow them to disambiguate their utterances in the most efficient manner.
To do so, this view lists decision variables with the highest information gain, determined by the algorithm introduced in  Section~\ref{sec:algorithm}. 
Even if decision variables yield high information gain, they may not necessarily capture the characteristics most relevant for disambiguation with respect to the user’s intent, since many intents are possible.
Thus, the user accesses one (grouped) decision variable at a time but can iterate through the alternatives by either clicking the up and down arrow keys on the keyboard or using the scroll function of the corresponding panel.
Underneath the red-marked decision variables, we display an example query that contains the particular variable.
This example should help contextualize and interpret 
the
given variable,
especially if it consists of multiple atomic features.
The query example is displayed using the global design rationale introduced in~ Section~\ref{sec:design-rationale}, i.e., it is visualized as a list of atomic features, whereby the decision variables are colored red. Additionally, we identify and highlight---using a low-opacity red overlay---the features in the example query that would be implicitly included in the final query if the user selects the decision variable (see, e.g., \verb|JOIN Film| ~feature in~Figure~\ref{fig:decision-predicted-query}, decision x+1).

By hovering over the decision variable, queries containing these features will be highlighted in the \textit{Action Space}.
By hovering over the example query, the particular query glyph will be highlighted. 
The most important interaction is the binary decision of whether to select a variable, thereby filtering for queries that include it, or to reject the variable, which instead filters queries that exclude it. 
The binary decision is made by clicking on the ``Yes'' or ``No'' button displayed under the query example. 
As shown in Figure~\ref{fig:decision}, by clicking on one of the buttons, queries that contain/exclude the variables are filtered in the \textit{Action Space}, and new decision variables are computed based on the new query sample.
In addition, the \textit{Predicted Query} panel (introduced in the next paragraph) are updated with atomic features that are likely to occur in the final query.
Every decision made can be reversed by clicking on the \textit{Back} button, as shown in~Figure~\ref{fig:decision-predicted-query}.
Alternatively, the users can simply skip the suggested decision variable and explore the alternative variables by using the up and down arrow keys or the scroll bar.

\paragraph{Predicted Query} 
The goal of the \textit{Predicted Query} is to support the provenance of the decisions made. The users are informed about their made decisions and the likely atomic features that will be included in the final query.
As shown in~Figure~\ref{fig:decision-predicted-query}, the atomic features are visualized in the same way as in the other two panels.
Unlike other panels, the atomic features that are considered likely but remain unconfirmed by the user are indicated with a surrounding border.
The likelihood of atomic features is computed based on the current query sample displayed in the \textit{Action Space}.
Every time a new query sample is created, by selecting queries in the \textit{Action Space} or decision variables in the \textit{Decision Space}, the atomic features are filtered (probability > 0) and their probabilities updated. 

In this panel, we apply a linking-and-brushing method similar to that in the \textit{Decision Space}. 
The user can hover over an atomic feature, and all queries containing the feature will be highlighted in the \textit{Action Space}.
Furthermore, the user can make the same binary decisions by clicking on ``Yes'' or ``No'' buttons to filter the \textit{Action Space}.
New decision variables will be computed based on the new query sample and listed in the \textit{Decision Space}, atomic features will be filtered, and their probabilities updated in this panel.

\section{User Study}

Through our user study, we examine how people use \textsc{PleaSQLarify} to disambiguate utterances and evaluate pragmatic repair in action for the application case of text-to-SQL. As outlined in Section~\ref{subsec:design-implications}, pragmatic repair involves helping users grasp and work within the available action space through concise, informative, and clear interactions.
We frame this objective around two key questions:

\begin{itemize}
\item \textbf{Understanding the Probable Action Space}: Does the system help users recognize and distinguish among the LLM’s possible interpretations of an utterance?
\item \textbf{Navigating the Action Space}: Do the decision variables enable users to move through this space in a way that feels efficient while remaining comprehensible?
\end{itemize}

Together, these aspects determine whether pragmatic repair successfully supports the process of ineractively grounding communicative intent in a complex technical domain.

\subsection{Participants}
We recruited 12 computer science graduate students and professionals for this study (age: $M = 27.3, SD = 1.8$; gender: 4F, 8M).
As users are expected to interpret and interact with LLM-generated SQL queries, we required each participant to have some basic familiarity with SQL.
Nine participants reported having some exposure to writing and interpreting SQL queries (intermediate), whereas three participants reported doing so regularly (advanced).
All participants reported having some experience using large language models for code generation or debugging.

\aptLtoX{\begin{table*}[t!]
\centering
\small
\caption{Task completion by participant.}
\Description{The table summarizes task completion by participant for five query tasks of varying ambiguity type and complexity. 
Each row lists a task, its ambiguity type, complexity level, and subtasks. Tasks include examples such as finding reviews of “Pulp Fiction,” identifying ratings of movies with Leonardo DiCaprio, and showing film categories offered by festivals. 
Participants P1 through P12 are shown across the top, with their SQL expertise level indicated as either intermediate or advanced. 
Cells within the table are color-coded to indicate completion outcomes: green for solved, yellow for partial success with a better query, orange for partial success without the original query, and red for not solved. 
For example, most participants solved the low-complexity “Pulp Fiction” and “Leonardo DiCaprio” tasks, while the high-complexity festival category task had many red cells, indicating it was not solved by most. 
The table also distinguishes subtasks 2a (identify interpretation clusters), 2b (understand semantic differences between clusters), and 3 (navigate to utterance interpretation). 
Overall, the table illustrates how task completion varied by task difficulty, subtask type, and participant expertise.}
\label{tab:task-completion}
\begin{tabular}{l l l l cccccccccccc}
\hline
\multicolumn{4}{c}{\textit{Participant ID}} & P1 & P2 & P3 & P4 & P5 & P6 & P7 & P8 & P9 & P10 & P11 & P12 \\
\multicolumn{4}{c}{\textit{SQL expertise}} & \A{A}  & \I{I}  & \I{I}  & \I{I}  & \A{A}  & \A{A}  & \I{I}  & \I{I}  & \I{I}  & \I{I}   & \I{I} & \I{I} \\ 
\hline
\textbf{Task} & Ambig. Type & Complexity & Subtask &  &&&&&&&&&&&\\
\hline
\multirow{3}{*}{\makecell[l]{\textbf{What was the} \break \textbf{review of}  \break \textbf{\textit{Pulp Fiction}?}}} 
    & \multirow{3}{*}{\makecell[l]{Column \break Vagueness}} & \multirow{3}{*}{Low}
    & 2a   & \yes & \yes & \yes & \yes & \yes & \yes & \yes & \yes & \yes & \yes & \yes & \yes \\
 & & & 2b & \yes & \yes & \yes & \yes & \yes & \yes & \yes & \yes & \yes & \yes & \yes & \yes \\
 & & & 3 & \yes & \yes & \yes & \yes & \partialmajor & \yes & \yes & \yes & \yes & \yes & \yes & \yes \\
\hline
\multirow{3}{*}{\makecell[l]{\textbf{What's the rating of the}  \break \textbf{movies with Leonardo}  \break \textbf{DiCaprio?}}} 
    & \multirow{3}{*}{\makecell[l]{Column \break Vagueness}} & \multirow{3}{*}{Medium}
    & 2a    & \yes & \yes & \yes & \yes & \yes & \yes & \yes & \yes & \yes & \yes & \yes & \yes \\
 & & & 2b  & \yes & \yes & \yes & \yes & \yes & \yes & \yes & \yes & \yes & \yes & \yes & \yes \\
 & & & 3 & \partialmajor & \partialminor & \yes & \yes & \yes & \partialminor & \partialmajor & \yes & \yes & \yes & \partialmajor & \partialmajor \\
\hline
\multirow{3}{*}{\makecell[l]{\textbf{Show me scripts and}  \break \textbf{editors with a deadline}  \break \textbf{of 21.04.03.}} }
    & \multirow{3}{*}{Attachment} & \multirow{3}{*}{High}
    & 2a   & \yes & \yes & \yes & \yes & \yes & \yes & \yes & \yes & \yes & \yes & \yes & \yes \\
 & & & 2b  & \yes & \no & \no & \yes & \no & \yes & \yes & \no & \no & \yes & \yes & \no \\
 & & & 3& \yes & \partialmajor & \partialmajor & \yes & \partialmajor & \yes & \partialmajor & \partialminor & \no & \yes & \yes & \yes \\
\hline
\multirow{3}{*}{\makecell[l]{\textbf{Show all the action}  \break \textbf{movies and romantic}  \break \textbf{comedies lasting 2 hours.}}} 
    & \multirow{3}{*}{Attachment} & \multirow{3}{*}{Medium}
    & 2a   & \yes & \yes & \yes & \yes & \yes & \yes & \yes & \yes & \yes & \yes & \yes & \yes \\
 & & & 2b  & \yes & \no & \no & \yes & \yes & \yes & \yes & \yes & \no & \yes & \yes & \yes \\
 & & & 3 & \yes & \yes & \no & \yes & \yes & \yes & \partialmajor & \partialminor & \yes & \yes & \partialminor & \yes \\
\hline
\multirow{3}{*}{\makecell[l]{\textbf{What film categories}  \break \textbf{does each film festival}  \break \textbf{offer?}}} 
    & \multirow{3}{*}{Scope} & \multirow{3}{*}{High}
    & 2a    & \yes & \yes & \no & \yes & \yes & \yes & \yes & \yes & \no & \yes & \yes & \no \\
 & & & 2b  & \no & \no & \no & \yes & \yes & \no & \no & \no & \no & \no & \no & \no \\
 & & & 3 & \yes & \no & \no & \yes & \yes & \yes & \partialminor & \yes & \no & \yes & \partialmajor & \no \\
\hline
\end{tabular}
\begin{tabular}{@{}l l@{\hspace{1.8cm}} l l@{\hspace{1.8cm}} l l@{}}
\multicolumn{2}{c}{Task completion} 
  & \multicolumn{2}{c}{SQL expertise} 
  & \multicolumn{2}{c}{Subtasks} \\[0.75ex]
\yes & solved 
  & \I{I} & Intermediate 
  & 2a  & Identify interpretation clusters \\
\partialminor & partial (found better query) 
  & \A{A} & Advanced  
  & 2b  & \makecell[l]{Understand semantic  \break difference between clusters} \\
\partialmajor & partial (didn't find original query) 
  & & 
  & 3 & Navigate to utterance interpretation \\
\no & not solved & & & &
\end{tabular}
\end{table*}}{\setlength{\tabcolsep}{4pt}
\renewcommand{\arraystretch}{1.4}
\begin{table*}[t!]
\centering
\small
\caption{Task completion by participant.}
\Description{The table summarizes task completion by participant for five query tasks of varying ambiguity type and complexity. 
Each row lists a task, its ambiguity type, complexity level, and subtasks. Tasks include examples such as finding reviews of “Pulp Fiction,” identifying ratings of movies with Leonardo DiCaprio, and showing film categories offered by festivals. 
Participants P1 through P12 are shown across the top, with their SQL expertise level indicated as either intermediate or advanced. 
Cells within the table are color-coded to indicate completion outcomes: green for solved, yellow for partial success with a better query, orange for partial success without the original query, and red for not solved. 
For example, most participants solved the low-complexity “Pulp Fiction” and “Leonardo DiCaprio” tasks, while the high-complexity festival category task had many red cells, indicating it was not solved by most. 
The table also distinguishes subtasks 2a (identify interpretation clusters), 2b (understand semantic differences between clusters), and 3 (navigate to utterance interpretation). 
Overall, the table illustrates how task completion varied by task difficulty, subtask type, and participant expertise.}
\label{tab:task-completion}
\begin{tabular}{l l l l cccccccccccc}
\toprule
\multicolumn{4}{c}{\textit{Participant ID}} & P1 & P2 & P3 & P4 & P5 & P6 & P7 & P8 & P9 & P10 & P11 & P12 \\
\multicolumn{4}{c}{\textit{SQL expertise}} &
\A  & \I  & \I  & \I  & \A  & \A  & \I  & \I  & \I  & \I   & \I & \I \\ 
\toprule
\textbf{Task} & Ambig. Type & Complexity & Subtask & \multicolumn{12}{c}{} \\
\midrule
\multirow{3}{*}{\makecell[l]{\textbf{What was the} \\ \textbf{review of} \\ \textbf{\textit{Pulp Fiction}?}}} 
    & \multirow{3}{*}{\makecell[l]{Column\\Vagueness}} & \multirow{3}{*}{Low}
    & 2a   & \yes & \yes & \yes & \yes & \yes & \yes & \yes & \yes & \yes & \yes & \yes & \yes \\
 & & & 2b & \yes & \yes & \yes & \yes & \yes & \yes & \yes & \yes & \yes & \yes & \yes & \yes \\
 & & & 3 & \yes & \yes & \yes & \yes & \partialmajor & \yes & \yes & \yes & \yes & \yes & \yes & \yes \\
\midrule
\multirow{3}{*}{\makecell[l]{\textbf{What's the rating of the} \\ \textbf{movies with Leonardo} \\ \textbf{DiCaprio?}}} 
    & \multirow{3}{*}{\makecell[l]{Column\\Vagueness}} & \multirow{3}{*}{Medium}
    & 2a    & \yes & \yes & \yes & \yes & \yes & \yes & \yes & \yes & \yes & \yes & \yes & \yes \\
 & & & 2b  & \yes & \yes & \yes & \yes & \yes & \yes & \yes & \yes & \yes & \yes & \yes & \yes \\
 & & & 3 & \partialmajor & \partialminor & \yes & \yes & \yes & \partialminor & \partialmajor & \yes & \yes & \yes & \partialmajor & \partialmajor \\
\midrule
\multirow{3}{*}{\makecell[l]{\textbf{Show me scripts and} \\ \textbf{editors with a deadline} \\ \textbf{of 21.04.03.}} }
    & \multirow{3}{*}{Attachment} & \multirow{3}{*}{High}
    & 2a   & \yes & \yes & \yes & \yes & \yes & \yes & \yes & \yes & \yes & \yes & \yes & \yes \\
 & & & 2b  & \yes & \no & \no & \yes & \no & \yes & \yes & \no & \no & \yes & \yes & \no \\
 & & & 3& \yes & \partialmajor & \partialmajor & \yes & \partialmajor & \yes & \partialmajor & \partialminor & \no & \yes & \yes & \yes \\
\midrule
\multirow{3}{*}{\makecell[l]{\textbf{Show all the action} \\ \textbf{movies and romantic} \\ \textbf{comedies lasting 2 hours.}}} 
    & \multirow{3}{*}{Attachment} & \multirow{3}{*}{Medium}
    & 2a   & \yes & \yes & \yes & \yes & \yes & \yes & \yes & \yes & \yes & \yes & \yes & \yes \\
 & & & 2b  & \yes & \no & \no & \yes & \yes & \yes & \yes & \yes & \no & \yes & \yes & \yes \\
 & & & 3 & \yes & \yes & \no & \yes & \yes & \yes & \partialmajor & \partialminor & \yes & \yes & \partialminor & \yes \\
\midrule
\multirow{3}{*}{\makecell[l]{\textbf{What film categories} \\ \textbf{does each film festival} \\ \textbf{offer?}}} 
    & \multirow{3}{*}{Scope} & \multirow{3}{*}{High}
    & 2a    & \yes & \yes & \no & \yes & \yes & \yes & \yes & \yes & \no & \yes & \yes & \no \\
 & & & 2b  & \no & \no & \no & \yes & \yes & \no & \no & \no & \no & \no & \no & \no \\
 & & & 3 & \yes & \no & \no & \yes & \yes & \yes & \partialminor & \yes & \no & \yes & \partialmajor & \no \\
\bottomrule
\end{tabular}
\vspace{0.5ex}
\begin{minipage}{\textwidth}\footnotesize
\begin{tabular}{@{}l l@{\hspace{1.8cm}} l l@{\hspace{1.8cm}} l l@{}}
\multicolumn{2}{c}{Task completion} 
  & \multicolumn{2}{c}{SQL expertise} 
  & \multicolumn{2}{c}{Subtasks} \\[0.75ex]
\yes & solved 
  & \I & Intermediate 
  & 2a  & Identify interpretation clusters \\
\partialminor & partial (found better query) 
  & \A & Advanced  
  & 2b  & \makecell[l]{Understand semantic \\ difference between clusters} \\
\partialmajor & partial (didn't find original query) 
  & & 
  & 3 & Navigate to utterance interpretation \\
\no & not solved & & & &
\end{tabular}
\end{minipage}
\vspace{-1em}
\end{table*}}

\subsection{Procedure}
Each user study lasted between 45 and 60 minutes and was composed of four main sections. Eight sessions were held in person, and four were held remotely due to limited participant availability.
For remote studies, participants were asked to share their screen while solving tasks.
Each session was recorded and transcribed for later analysis.
Informed consent was obtained from each participant, and participants were compensated with 25\$.

\paragraph{Introduction (5 Minutes)} After welcoming the participants, we gave them a brief overview of the study procedure and, if required, helped them set up \textsc{PleaSQLarify} on their device. We then collected participants' demographic information along with their SQL expertise and experience using LLMs for coding. 

\paragraph{System Onboarding (15 Minutes)} To familiarize participants with the system features, we let them watch a 5-minute system introduction video, which walked through all relevant system components. We then let the users test those features on a test sample, which was not part of the ones shown in the task. Namely, we made sure participants knew how to access the test database, interact with datapoints in the \textit{Action Space}, and navigate using the \textit{Decision Space} and the clauses in the \textit{Predicted Query} panel.

\paragraph{Tasks (25 Minutes)}
Once the participants declared to feel comfortable using the system's features, they were asked to complete 5 tasks of varying difficulty using \textsc{PleaSQLarify}.
To understand participants' behavior and reasoning during task-solving, they were asked to think aloud.
When solving the tasks, participants were free to ask questions at any time. If a participant's answer to a task was unclear, the conducting authors would ask the participant to clarify.\looseness=-1

\paragraph{Feedback (10 Minutes)} Finally, participants were asked to answer questions about the system's overall utility compared to standard language model querying and provide feedback about each component. Notably, this included a comparison between grouped, prioritized decision variables in the \textit{Decision Space} and the baseline of ordered, atomic decision variables in the \textit{Predicted Query} panel.
Additionally, participants completed a system usability score (SUS) questionnaire to evaluate the usability of the system. During and after filling out the questionnaires, participants could verbalize any additional qualitative feedback about their experience with the system.\looseness=-1

\subsection{Tasks}

\paragraph{Data.} We design five tasks using AMBROSIA samples drawn from our quantitative evaluation in  Section~\ref{sec:evaluation}.
Each AMBROSIA utterance has at least two high-level semantic interpretations, for which the corresponding gold SQL queries are provided.
Each AMBROSIA sample contains a small database of around 5 tables, each with <~10 rows and columns per table.  We chose all samples from the same domain (Filmmaking) to reduce cognitive load. We selected two samples from each of the ``vague'' and ``attachment'', and one from the ``scope'' ambiguity type.
Whereas the ``vague'' queries mainly contained rather basic column ambiguities induced by imprecise wording, the ``attachment'' and ``scope'' types were more challenging---both in the reasoning needed for disambiguation and in the length and complexity of the resulting SQL programs. We provide an overview of the study tasks along with the users' task completion in Table \ref{tab:task-completion}.

\begin{figure*}[t!]
    \centering
    \includegraphics[width=\linewidth]{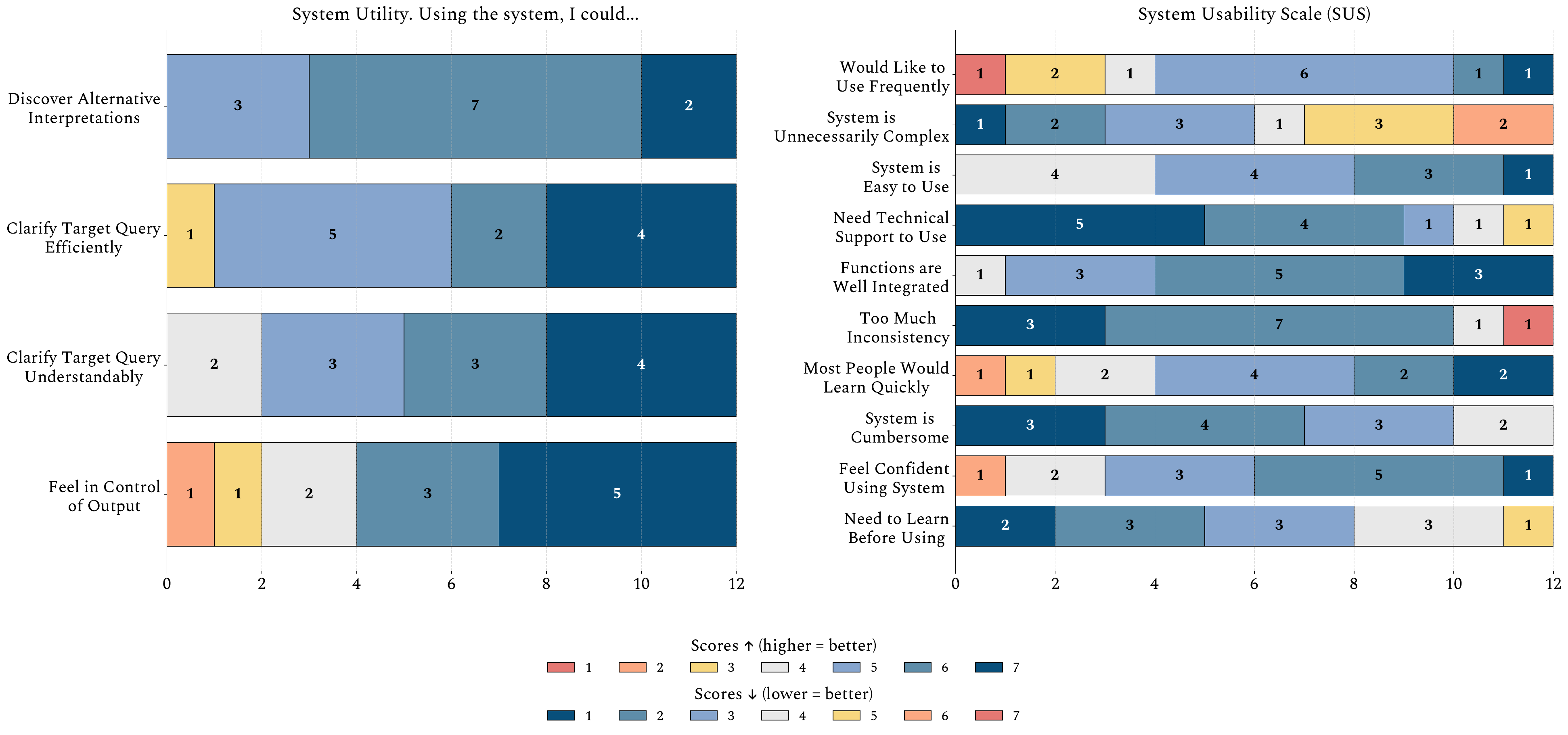}
    \caption{System utility and usability questionnaire responses.}
    \Description{Bar charts summarizing questionnaire responses. 
    On the left, system utility ratings show participants could discover alternative interpretations, clarify queries efficiently and understandably, and feel in control of outputs. On the right, System Usability Scale (SUS) items show mixed but generally positive responses, with most users finding the system easy to use, functions well integrated, and expressing confidence, though some noted complexity and inconsistency.}
    \label{fig:sys-utility-sus}
\end{figure*}

\paragraph{Subtasks.} In alignment with the user study's objectives, we split the task into three parts: 
(1) \textit{Utterance interpretation}: Given an AMBROSIA ambiguous utterance and the database, each participant was asked to verbally describe a plausible interpretation or output. 
(2) \textit{Exploration}:  Participants then examined the candidates and their clustering
to understand both a) functional differences (i.e., how the outputs varied) and b) structural differences (i.e., which SQL clauses accounted for those differences) among them.
(3) \textit{Navigation}:  Finally, participants were tasked with narrowing the candidate set down to the one that best matched their original \textit{utterance interpretation}.

\section{Results}

\subsection{General Remarks}

\paragraph{Task Completion. } Table \ref{tab:task-completion} shows that, using our system, participants achieved an 84.4\% overall task completion rate\footnote{Average completion rate across all 180 tasks, counting partial solutions as correct.} across varying SQL expertise levels. Importantly, every task was completed at least once. Subtask 2a required describing the samples in the \textit{Action Space} clusters. Most participants solved it across tasks, except for P3, P9, and P12, who chose to drop out for time reasons after completing task 4. Subtask 2b asked participants to explain the semantic differences between clusters. This was especially difficult in task 3, where the difference was due to a subtle filtering condition, and in task 5, which involved many joins and a complex filtering requirement.\footnote{One implementation required checking whether the category was present at \textit{every} film festival, requiring an aggregation logic to count and filter festival occurrences.} For subtask 3, around half of the participants (7/12) struggled with not finding their original utterance interpretation for at least one of the tasks. However, when prompted to consider alternatives, they were able to arrive at different satisfactory final interpretations, as shown in orange in Table \ref{tab:task-completion}.

\paragraph{System Usability.} The SUS results in Figure~\ref{fig:sys-utility-sus} indicate that participants found the system overall enjoyable and relatively easy to use ($M=5.08$ on a 7-point Likert scale) with well-integrated components ($M=5.83$). In line with this, participants commented that the interface was ``\textit{very nice to use}'' (P4) with a ``\textit{smooth}'' (P2) workflow. Most participants thought that they could interact with it confidently after only a short period of exploration ($M=5.17$). 
Whereas a few participants noted an initial learning curve (P6, P7, P11), they described it as short-lived and not diminishing the overall impression of usability.
The positive SUS ratings suggest that \textsc{PleaSQLarify} not only supports task solving but also provides an interaction model that participants would be willing to adopt, for instance, to identify unexpected behavior (P6) or to support query-writing with increased user control (P10).

\subsection{Understanding the Probable Action Space}

\paragraph{Discovering Alternative Interpretations.} Figure~\ref{fig:sys-utility-sus} shows that participants rated the system highly on helping them recognize and explore multiple interpretations of an utterance ($M=5.92$ on a 7-point Likert scale) compared to conventional language model querying. This suggests that, in contrast to the single-response paradigm of traditional LLM querying, the \textit{Action Space} view and clustering helped convey the breadth of possible LLM interpretations. Indeed, several participants (P1, P2, P6,  P10) explicitly reported that they had not recognized the ambiguity of the original utterance until they were presented with alternative interpretations.
For instance, when P6 explored the attachment ambiguity sample 4, they noticed additional vagueness in the query; ``\textit{Oh, `lasting two hours', does this mean exactly two hours or longer than two hours? I haven't actually thought about that.}'' This indicates that surfacing the action space supports the first design requirement: making the LLM’s action space visible in a way that broadens user awareness. 

\paragraph{Discovering Better Alternatives.} Almost half of the participants (5/12) reported discovering alternatives they considered more useful than their originally intended query, even when the latter was included in the candidate set. We show these instances in yellow in Table \ref{tab:task-completion}.
For example, during task 2, P6 realized it would be helpful to include the title column after inspecting the current output.
Similarly, while finalizing their selection, P5 noticed that a grouping would look nicer, and P6 noticed in the final step of task 3 that they could additionally order the table for readability. 

\paragraph{Query-Writing Guidance. } P2 and P10 reported that the output-based approach, coupled with the decision variables, helped them write queries that they wouldn't have known to write on their own. This may be due to the fact that the system breaks down complex operations into digestible decisions that operate at an understandable level. 

\subsection{Navigating the Decision Space}

\paragraph{Perceived Control. }
By converting LLM generation under uncertainty into an explicit decision process, the system made participants feel a greater control over the generated output ($M=5.50$ on a 7-point Likert scale, cf. Figure~\ref{fig:sys-utility-sus}) compared to traditional language model querying.
This was reflected in the fact that most participants were able to reach their desired output using the system (cf. Table \ref{tab:task-completion}). Some, however, felt that they were at times being too strongly guided by the decision algorithm (P3, P7). For instance, P3 felt like they were ``\textit{validating more, rather than being in control}''. Others felt overly constrained by the original candidate set (P2, P5, P7, P11). 
For example, P5 consistently wanted to view all columns to get an overview, but this option was rarely included in the candidate set. Similarly, in task 3, P7 remarked: ``\textit{I was expecting a [few] more columns; I would want to show everything about the [table]}''. We revisit these concerns in the limitations (cf.  Section~\ref{sec:limitations}). 

\begin{figure*}[t!]
    \centering
    \includegraphics[width=0.95\linewidth]{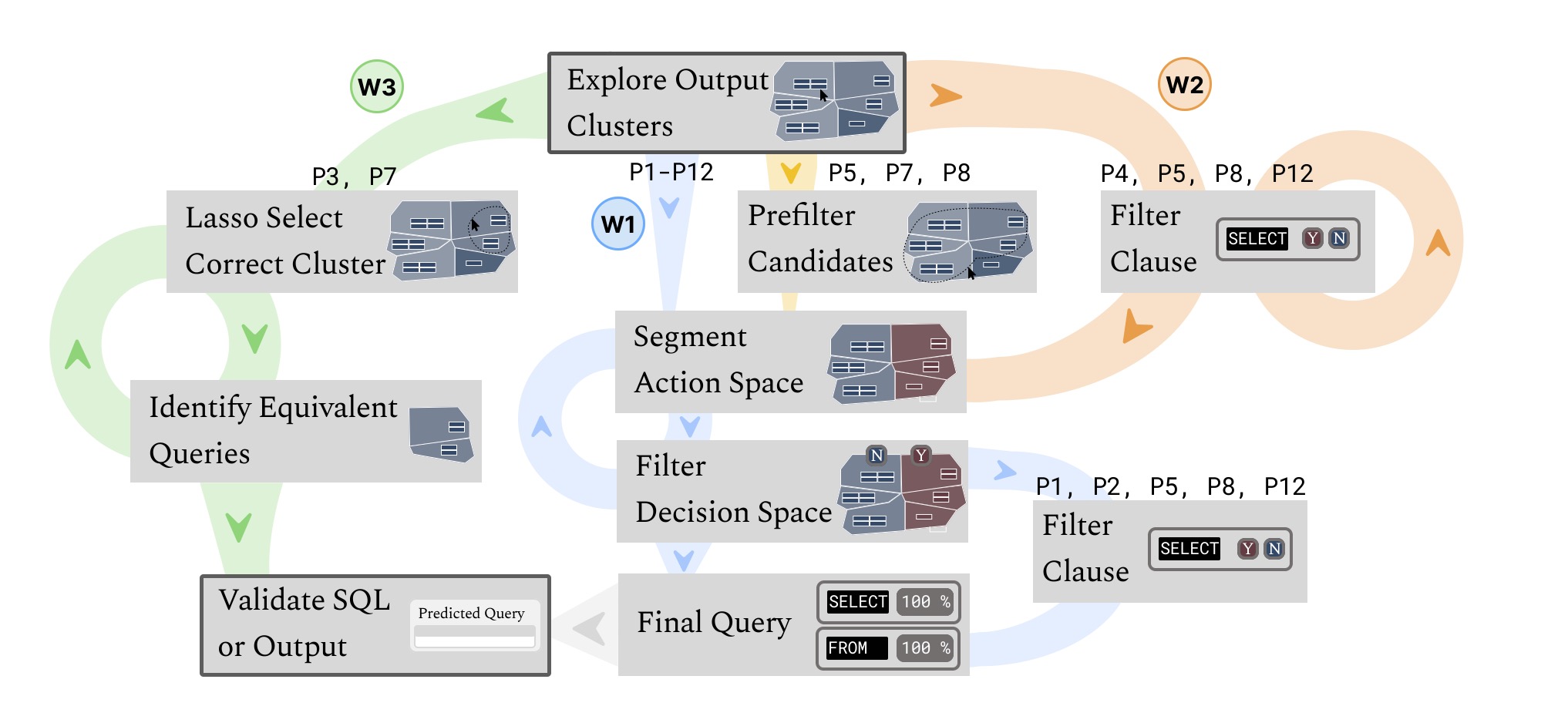}
    \vspace{-1em}
    \caption[]{Observed user interaction flows during exploration and intent clarification. After exploring the initial action space, three common approaches for intent clarification were chosen. In workflow~\protect\aptLtoX{\raisebox{-3pt}{\includegraphics[width=0.022\textwidth]{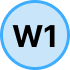}}}{\protect\inlinegraphics{figures/w1.png}}, users navigated to their target query primarily using the \textit{Decision Space}, whereas in~\protect\aptLtoX{\raisebox{-3pt}{\includegraphics[width=0.022\textwidth]{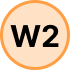}}}{\protect\inlinegraphics{figures/w2.png}} users navigated by filtering individual clauses in the \textit{Predicted Query} panel. Finally, an output-based approach emerged~\protect\aptLtoX{\raisebox{-3pt}{\includegraphics[width=0.022\textwidth]{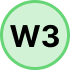}}}{\protect\inlinegraphics{figures/w3.png}} in which users explicitly chose desired output behavior.}
    \Description{Diagram of observed user interaction flows for intent clarification. 
    Three workflows are shown: W1, where users segment and filter the Decision Space to reach a final query; W2, where users filter individual clauses in the Predicted Query panel; 
    and W3, an output-based approach where users select clusters, identify equivalent queries, 
    and validate SQL or outputs. Arrows connect the steps, illustrating alternative navigation paths.}
    \vspace{-1em}
    \label{fig:workflows}
\end{figure*}

 \paragraph{Interaction Efficiency.} Most participants rated the system positively for supporting \textit{efficient} clarification of their queries ($M=5.67$ on a 7-point Likert scale, cf. Figure~\ref{fig:sys-utility-sus}). Participants attributed this primarily to the decision variable prioritization algorithm in the \textit{Decision Space}: instead of rewriting queries, they could incrementally guide the system and often reach a satisfactory interpretation within a few steps.
Corroborating this, participants found their interaction with the prioritized, grouped decision variables in the \emph{Decision Space} panel less effortful than with the atomic decision variables in the \emph{Predicted Query} panel (grouped: $M = 3.25$ vs. atomic: $M = 4.42$).
 For example, P3 likened the prioritization-based decision process to  ``\textit{binary search on the query, basically depending on the cluster sizes}.'' Participants further reported that the initially surfaced decision variables were typically perceived as more relevant, whereas later variables were described as ``\textit{less relevant}'' (P6) or ``\textit{optional}'' (P7). This highlights the utility of our algorithm for surfacing semantically important features. 

\paragraph{Decision Understandability.} The system also received positive ratings for making decisions \textit{understandable} ($M=5.75$ on a 7-point Likert scale, cf. Figure~\ref{fig:sys-utility-sus}). Although grouped and atomic decision variables were not always easy to interpret in isolation (grouped: $M=5.33$, atomic: $M=4.92$), participants could contextualize them using the example query and their implicit decisions in the \textit{Decision Space} panel. 
Some participants noted that prioritization occasionally made choices feel less structured and made it harder to track progress in the decision process (P1, P2, P5, P8, P12).
To this end, toward the end of the clarification process, participants would glance at or rely on the \textit{Predicted Query} panel for orientation.

\begin{figure*}[t!]
    \centering
    \includegraphics[width=\linewidth]{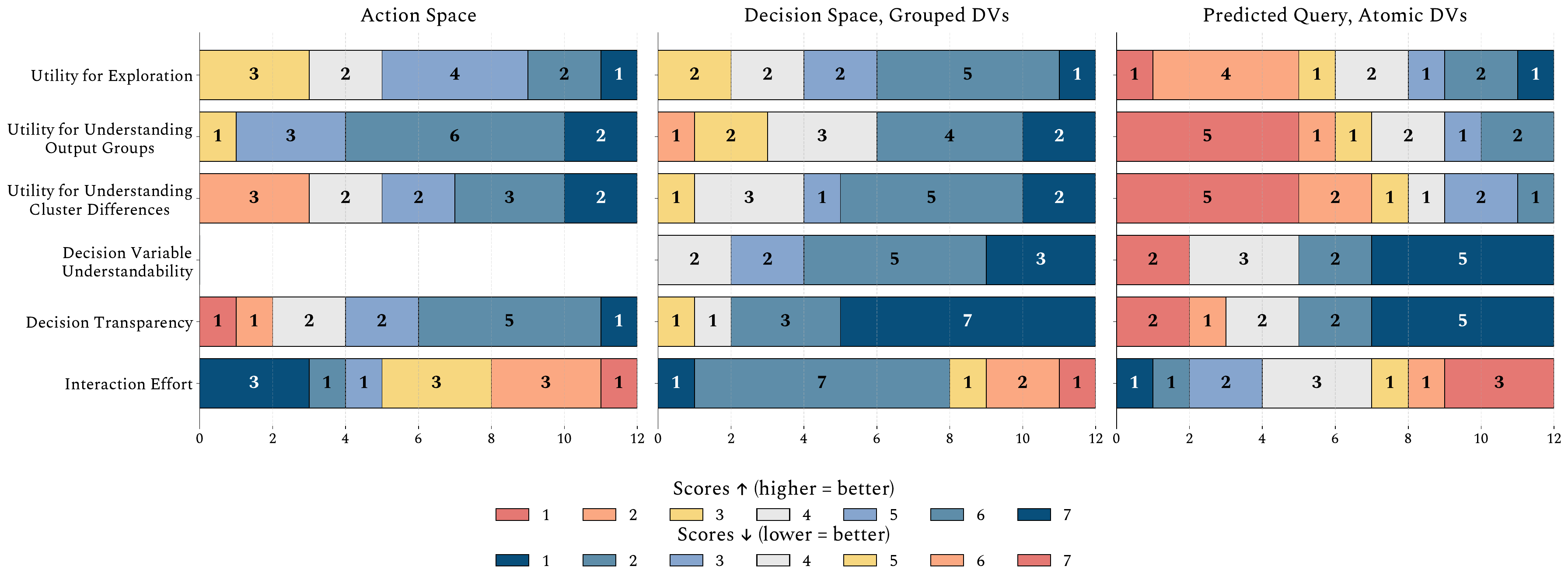}
    \caption{Component utility questionnaire responses, grouped by component.}
    \Description{Stacked bar charts of component utility questionnaire responses across three interface components: Action Space, Decision Space with grouped decision variables, and Predicted Query with atomic decision variables. Ratings cover utility for exploration, understanding outputs and clusters, decision variable understandability, decision transparency, and interaction effort. Action Space and grouped Decision Space are rated more positively, while atomic Predicted Query shows lower scores and higher effort.}
    \vspace{-1.5em}
        \label{fig:components}
\end{figure*}

\subsection{Usage Patterns and System Component Analysis}\label{subsec:usage-patterns}
As the system provides multiple ways to explore and navigate the candidate set, participants employed different approaches while solving tasks.
We illustrate these workflows in Figure~\ref{fig:workflows} and describe them below.

\paragraph{1. Initial Exploration.} Exploration in \textsc{PleaSQLarify} consistently began in the \textit{Action Space}, which provided an overview of the candidate query landscape and allowed participants to orient themselves. 
They scanned clusters and hovered over glyphs to inspect tooltips. Some participants started by excluding candidates they recognized as implausible or “\textit{hallucinated}” (P5, P7, P8), while others (P3, P7) used lasso selection to focus on relevant clusters and generate decision variable splits. This universal entry point set the stage for three distinct workflows, each emphasizing one interface component.

\paragraph{2a. Decision-Based Navigation. } Workflow~\aptLtoX{\raisebox{-3pt}{\includegraphics[width=0.022\textwidth]{figures/w1.png}}}{\inlinegraphics{figures/w1.png}} was most common. In this workflow, participants first investigated clusters in the \textit{Action Space} and then used the \textit{Decision Space} to filter by grouped and ranked decision variables until converging on the intended query. Participants mainly relied on decision variables, while selecting simple clauses in the \textit{Predicted Query} panel (P1, P2, P5, P8, P12).
\aptLtoX{\raisebox{-3pt}{\includegraphics[width=0.022\textwidth]{figures/w1.png}}}{\inlinegraphics{figures/w1.png}} illustrates how the \textit{Action Space} and \textit{Decision Space} complemented one another: the \textit{Action Space} was rated more useful for contrasting functional output groups ($M=5.67$ vs. $4.83$, cf. Figure~\ref{fig:components}), whereas decision variables were rated more useful for understanding those differences ($M=5.33$ vs. $4.67$, cf. Figure~\ref{fig:components}). Participants typically hovered over the \textit{Predicted Query} panel to track their progress, using clause selection mainly at the start (e.g., fixing \verb|SELECT| clauses) or near the end when few decisions remained. Although some participants noted an initial learning curve (P6, P8), many settled on this workflow because they found it most efficient for identifying semantically meaningful candidate groups (P4, P7, P10).

\paragraph{2b. Clause-Level Navigation. }
Workflow~\aptLtoX{\raisebox{-3pt}{\includegraphics[width=0.022\textwidth]{figures/w2.png}}}{\inlinegraphics{figures/w2.png}} relied heavily on atomic clause selection in the \textit{Predicted Query} panel and was typically used when participants already knew the exact query they were aiming for (P5, P12). Instead of relying on decision variables, these users filtered the candidate set by selecting clauses directly and consulting tooltips. As P4 explained, “\textit{if you know exactly what you are looking for, the tooltip and selecting clauses is useful},” while P5 described this workflow as the fastest way to reach their intended result.
Many participants rated clause-level selection as more understandable than grouped decision variables due to the latter's variable prioritization (cf. Figure~\ref{fig:components}), yet also reported that it required significantly more effort due to the need to scroll through long lists without prioritization ($M=4.42$ vs. $3.25$, cf. Figure~\ref{fig:components}). Consequently, most participants preferred the \textit{Decision Space} for complex queries.
For example, P4 and P11 noted that while they used clause filtering at the start, they switched to prioritized decision variables once navigation became more complex and lists of possible \verb|WHERE| clauses became difficult to distinguish.

\paragraph{2c. Output-Based Navigation. } Workflow~\aptLtoX{\raisebox{-3pt}{\includegraphics[width=0.022\textwidth]{figures/w3.png}}}{\inlinegraphics{figures/w3.png}} was based on selecting promising outputs directly in the \textit{Action Space}. It was mainly adopted by participants (P3, P7) who had an expectation of the output shape but were less certain about the corresponding clauses. These users relied on lasso selection to isolate clusters, compared equivalent queries, and validated results in the \textit{Predicted Output} view. Although this approach required prior knowledge of the desired output structure, it naturally marginalized over equivalent programs, which participants found harder to distinguish through clause-level navigation ($M=4.67$ vs. $2.67$, cf. Figure~\ref{fig:components}). This workflow became more common as query complexity increased. For instance, P2 reported that for simple column ambiguities in early tasks, clause-level filtering was sufficient, but for more complex queries, they increasingly relied on \textit{Action Space} segmentation and output comparison, as clauses became too complex to interpret. 

\paragraph{3. Verification. } Regardless of the chosen workflow, participants concluded their analysis by validating the final query in the \textit{Predicted Query} and \textit{Predicted Output} panels.

\section{Discussion}

\subsection{General Design Implications}
Our study of \textsc{PleaSQLarify} not only demonstrates its effectiveness for text-to-SQL repair but also points to broader lessons for the design of interactive natural language interfaces. The patterns we observed suggest several principles for how interfaces can better support users in clarifying underspecified intent and resolving ambiguity.
\begin{enumerate}
  \item \textbf{Understanding ambiguity.} In structured natural language tasks such as text-to-SQL, ambiguity is often multi-layered, with several distinct interpretations coexisting without the user’s awareness. As a result, resolving such ambiguity is a decision process that unfolds over multiple interaction turns. Participants could consistently align the system with their intent when the interface surfaced this broader interpretation space, allowing them to recognize alternative meanings of their utterances and compare their outcomes.
    \item \textbf{Clarifying pragmatically.} Participants could steer the system effectively when clarification steps reduced the candidate space in ways that were both efficient and interpretable. Efficiency was supported by prioritizing decision variables, whereas interpretability emerged when decisions were contextualized: participants understood choices best when they were grounded in concrete examples, linked to their role in the query, or surfaced as implicit choices that the system made on their behalf.
  \item \textbf{Accommodating heterogeneous repair strategies.} The three workflows discussed in Section~\ref{subsec:usage-patterns} the complementary strengths of the \textit{Action Space}, \textit{Decision Space}, and \textit{Predicted Query}, as well as the flexibility with which participants appropriated them. The \textit{Action Space} supported broad exploration, the \textit{Decision Space} provided structured and prioritized guidance, and the \textit{Predicted Query} helped participants monitor convergence. Rather than enforcing a single path, users of future systems may benefit from having different strategies available---from open-ended clarification, to efficient targeting, to output-driven reasoning---depending on user goals and task complexity.
\end{enumerate}

More broadly, pragmatic repair extends beyond SQL.
The challenge of misaligned priors outlined in Section~\ref{subsec:pragmatics} and the framework of pragmatic repair presented in Section~\ref{subsec:repair} apply to most natural language interfaces.
By combining algorithmic selection of informative clarifications with visual techniques that reveal alternatives and make consequences observable, systems can leverage ambiguity as a resource for collaborative problem solving---supporting user control even when model interpretations misalign with user intent.\looseness=-1

\subsection{Limitations and Future Work}\label{sec:limitations}

\paragraph{Experimental Scope. } As the scope of the study is about intent clarification, the experimental setup (cf.  Section~\ref{sec:algorithm}) assumes the user has a well-formed intent that can be reasonably mapped to a valid SQL query. Other interesting related scenarios that were observed during the user study---such as when a user is unsure about their intent or changes their intent---are not covered by the experiments.

\paragraph{Limited Candidate Pool. } To control latency and computational cost, the candidate set was fixed to the initial set of around 50 language model generations. Although this may be enough to give a user an insight into the model's probable action space, it failed to adapt more dynamically to user requests. As a result, some participants reported not being able to find their exact intended query (P5, P7, P9). An interesting extension may be to update the candidate pool dynamically, given the user's input; however, this is not a feasible option with the current setup. As shown in Figure~\ref{fig:entropies_sims}, when evaluated on the AMBROSIA dataset, our algorithm---with optimal clarification behavior\footnote{That is, without selecting the wrong decision variable value.} and a fixed candidate pool---takes two to eight clarification steps until converging on the target query. Resampling 50 model generations at each step would therefore require up to 400 API calls for a single clarification sequence, introducing substantial latency and disproportionate computational costs.
Future work could address this limitation through adaptive resampling strategies, such as regenerating fewer samples at later stages when expected diversity decreases, or through an on-demand approach where users can choose to regenerate the candidate set if they encounter an underrepresented region of the action space.

\paragraph{Usability for Non-Technical Users.} While user study participants varied in SQL expertise, the system requires basic SQL knowledge to disambiguate queries, as users needed to interpret clauses and assess candidate outputs. Although visual, output-driven workflows were reported to improve accessibility to hard-to-write queries, our decision variables may still be hard to understand for non-SQL-literate users. An integration of code-to-text approaches may help further lower this bar \cite{narechania2021diy, liu2023abstraction}.

\paragraph{End-User Integration.} The visual interface presented in  Section~\ref{sec:design} is primarily a research tool to track and analyze user workflows when disambiguating utterances for text-to-SQL, rather than for use by end-users. However, individual elements, such as extracting decision variables or compact action space overviews, could be adapted into lightweight widgets and reused in conversational interfaces.\looseness=-1

\paragraph{Scalability.} The current setup explores small test databases as frequently used in the most popular text-to-SQL benchmarks \cite{yu2018spider, saparina2024ambrosia}.
Limiting the database size in this way enables a controlled quantitative evaluation as  Section~\ref{sec:evaluation}, as we can easily enumerate possible gold label intents, and it allowed participants in our user study to familiarize themselves with the database within a reasonable time frame.
For larger databases, our approach would likely take more clarification steps to converge, as the increased number of tables and columns expands the space of plausible query interpretations. Additionally, our proposed output-based analysis may become quite costly and harder to visualize.
Interesting future work may generate and perform analysis on smaller, informative subsets of larger databases to validate the utility of output-based distinctions between queries in more realistic setups.

\section{Conclusion}

We presented \textsc{PleaSQLarify}, a system that operationalizes pragmatic repair for interactive text-to-SQL disambiguation. The system combines an algorithm that selects informative decision variables with a visual interface that surfaces the model’s action space, requests incremental user clarifications, and makes updates traceable across turns. In a study with twelve participants, it helped users discover interpretations they had not anticipated and feel more in control of the generated outputs by efficiently and understandably guiding them toward satisfactory results.
The diversity of usage workflows further highlights the importance of supporting heterogeneous repair strategies.
More broadly, this work demonstrates how ambiguity in natural language interfaces can be turned into a resource for collaborative problem solving, strengthening user control even when model interpretations misalign with user intent.\looseness=-1

\begin{acks}
    The authors thank the anonymous reviewers for their thorough and constructive feedback. 
    The authors would also like to thank Ben Lipkin, Emily Cheng, Raphaël Baur, and Yanis Schmit for helpful discussions and feedback.
    Finally, the authors acknowledge funding from the SNF grant no. 10003068 and foryouandyourcustomers.\looseness=-1
\end{acks}

\bibliographystyle{ACM-Reference-Format}
\bibliography{sample-base}



\end{document}